\documentclass[12pt]{iopart}

\usepackage{graphicx}
\usepackage{wrapfig}

\usepackage{cite}

\newcommand{\be}{\begin{equation}}
\newcommand{\ee}{\end{equation}}
\newcommand{\bea}{\begin{eqnarray}}
\newcommand{\eea}{\end{eqnarray}}

\begin{document}

\title[Studies of superconductivity of Fe chalcogenides in films grown by PLD technique]{Studies of superconductivity of Fe chalcogenides in films grown by PLD technique}

\author{Atsutaka MAEDA$^{1,2}$ Tomoki KOBAYASHI$^{3}$, and Fuyuki NABESHIMA$^{4}$}

\address{{}$^1$ Department of Basic Science, University of Tokyo, 3-8-1, Komaba, Meguro-ku, Tokyo, 153-8902, Japan\\
{}$^2$ Physical Society of Japan, 2-31-22-5F, Yushima, Bunkyo-ku, Tokyo, 113-0034, Japan, \\
{}$^3$ Department of Physics, Osaka University, 1-1 Machikaneyama, Toyonaka, Osaka 560-0043, Japan\\
{}$^4$ JEOL, 3-1-2, Musashino, Akishima, Tokyo, 196-8558, Japan
}
\ead{cmaeda@g.ecc.u-tokyo.ac.jp}
\vspace{10pt}
\begin{indented}
\item[]August 2025
\end{indented}

\begin{abstract}
Studies on Fe chalcogenide superconductor using thin films grown by the PLD technique are reviewed in terms of electronic phase diagram, properties in the normal state, properties in the superconducting state, together with the comparison with properties in bulk crystals, MBE grown films and exfoliated crystals.
Challenges to increase superconducting $T_c$ will also be introduced.
\end{abstract}

%
%
%
%
%

\section{Introduction: Fe chalcogenides -valuable playgrounds for superconductivity of Fe based
materials-}

\begin{figure}[htb]
\begin{center}
\includegraphics[scale=0.25]{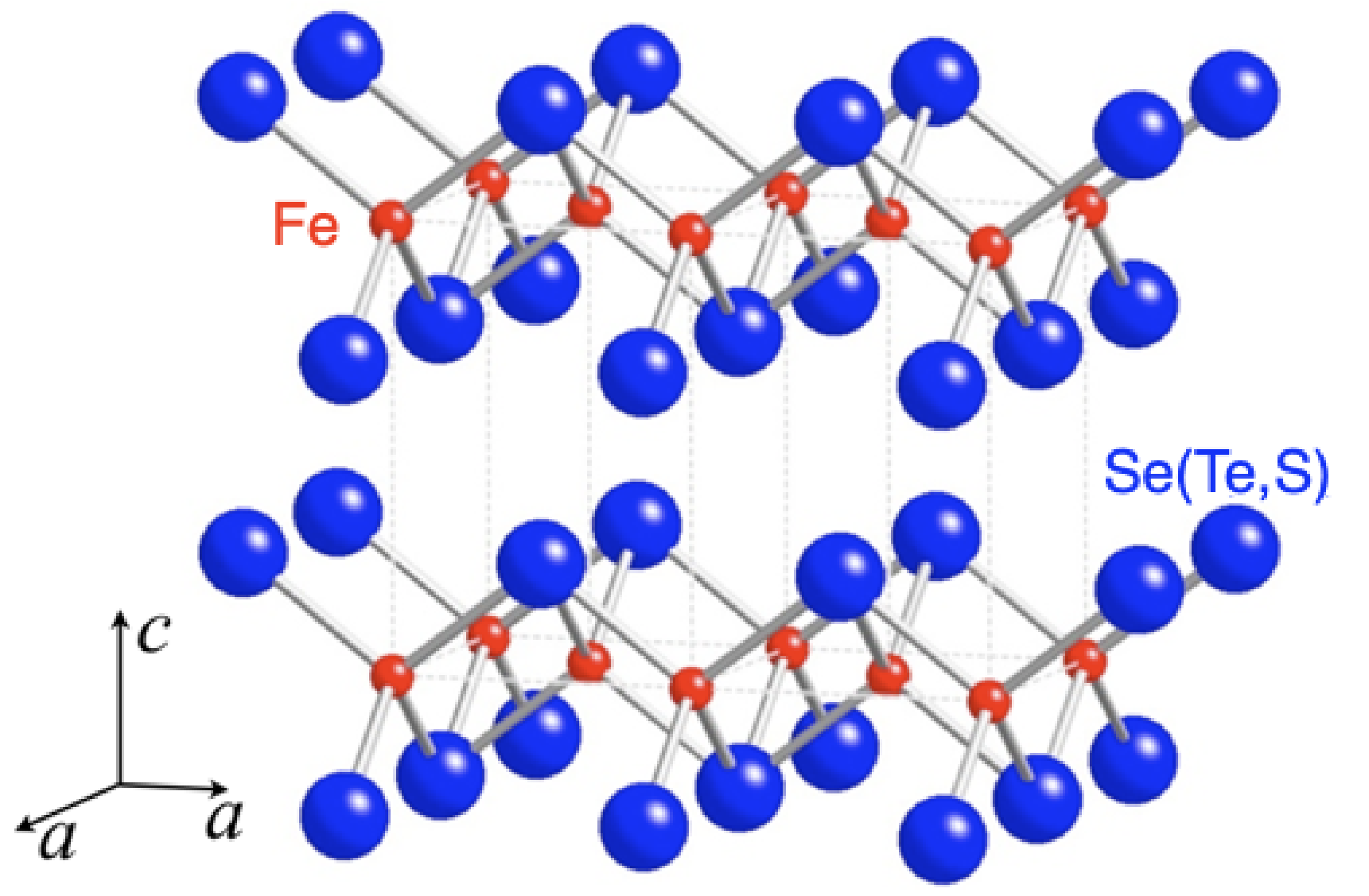}
\caption{
The crystal structure of FeSe$_{1-x}$S$_x$ and FeSe$_{1-y}$S$_y$.}
\label{crystal}
\end{center}
\end{figure}

Iron chalcogenide (FeCh) FeSe (Fig.~\ref{crystal}) occupies a unique position in the iron-based superconductor family.
Although it is a superconductor with $T_c \simeq $ 8 K under ambient pressure\cite{Hsu}, $T_c$ can be increased to 40-50 K by pressure application\cite{Mizuguchi,Medvedev,Miyoshi,Kothapalli,JPSun}, chemical substitution (including intercalation)\cite{Guo,Miyata,CHPWen,Ying,Lucas,Hosono}, and carrier doping such as field-effect doping\cite{Shiogai,Hanzawa,Lei}.
It has also been reported that when a monolayer film is fabricated on a certain oxide substrate, $T_c$ can reach 65 K or higher\cite{QYWang,SHe}.
This can be said as follows.
It has a unique characteristic that three different categories of superconductivity (SC) appear in the same material.
One is SC with the highest $T_c$ of about 40 K with electron Fermi surfaces (FSs) at the M point and hole FSs at the $\Gamma$ point\cite{XLiu,Bohmer,Coldea,Kreisel,Shibauchi}.
The second is SC of 40 - 50 K $T_c$ with only electron FSs at the M point, which is realized by intercalation or field-effect doping.
The third is possible high-temperature SC realized in ultrathin films on oxide substrates, which has been reported to have a $T_c$ of over 65 K.

Naturally, this third category of SC has attracted the interest of many researchers and has been actively studied.
Until now
Almost all the category 3 high-$T_c$ SC researches have been conducted using the molecular beam epitaxy (MBE) technique\cite{LWang,ZWang,DHuang,CLiu2020,SHan2021,TTanaka2021}.
The electronic structure is thought to be almost the same as that of category 2\cite{DLiu}, but some interface effects are believed to contribute to the high $T_c$.
Specifically, they may be due to interactions with phonon in the substrate\cite{JLee,YCui2015,CTang2016,WZhang2016,SZhang2016,SRebee2017}, electron doping from the substrate\cite{DLiu,WZhang2017,GZhou2018,WZhao2018}, or strain effects.
 In recent years, due to the remarkable progress in both sample preparation and measurement techniques, experimental studies of two-dimensional superconductors at the atomic layer level have become very active\cite{Saito}.
However, in most cases, the $T_c$ of superconductivity is the same as or lower than that of the bulk (for example, in the case of ultrathin film superconductors fabricated on semiconductor substrates,\cite{TZhang,Uchihashi,MYamada}, etc.).
In that sense, the large increase in $T_c$ of FeSe due to ultrathin film formation has stimulated many researchers, and in some cases, it is expected that this superconductivity may be the first realization of the so-called exciton mechanism\cite{Ginzburg,Bardeen}, which has been theoretically proposed for about half a century.
Now, the interface superconductivity of FeCh are reported in almost 10 kinds of substrate\cite{SRebee2017,BZhao2024,WZhang2014,diamag2,GZou2016,PZhang2016,HDing2016,RPeng2014,HYang2019,YSong2021,CLiu2021,CLi2024,GZhou2021}

However, with the exception of diamagnetism reports\cite{diamag2,YSong2021} and $\mu-$SR data\cite{Biswas2018}, all from a Fudan group, most of high-$T_c$ superconductivity reports have been spectroscopic, using STM\cite{QYWang} and angle-resolved photoemission spectroscopy (ARPES)\cite{SHe}, and the onset $T_c$ and the zero resistance temperature are at most 45-50 K and 30 K, respectively\cite{QYWang,Pedersen,Faeth,BZhao2024}.
Even in the most recent reference\cite{BZhao2024} where the authors got a rather good agreement between the temperature dependence of STS spectrum and that of dc resistivity, $T_c^{onset}$ is about 50 K and $T_c^{zero}$ is 30 K.
Thus, 65 K superconductivity has not appeared in the transport measurement yet.
In that sense, even now, 13 years after the first report, the enhancement of superconductivity has not been confirmed by electrical resistance.
There is a singular report that zero resistance was observed at 100 K by {\it in situ} measurement\cite{Ge}, but there have been no other reports of reproducibility, and doubts have been cast on its credibility\cite{Bozovic}.

The superconducting mechanisms of each category and their relationships are still important subjects of debate.
For category 1, researches using bulk single crystals are progressing\cite{XLiu,Bohmer,Coldea,Kreisel,Shibauchi}, and it has been proposed that the SC has been caused by nematic fluctuations, and /or magnetic fluctuations.
Moderately strong electron correlation leads to an anisotropic superconducting gap despite a very small FS in some materials, and orbital-selective pairing was proposed\cite{Sprau}.
In general, multi-band features show up in many physical properties.

\begin{figure}[htb]
\begin{center}
\includegraphics[scale=0.5]{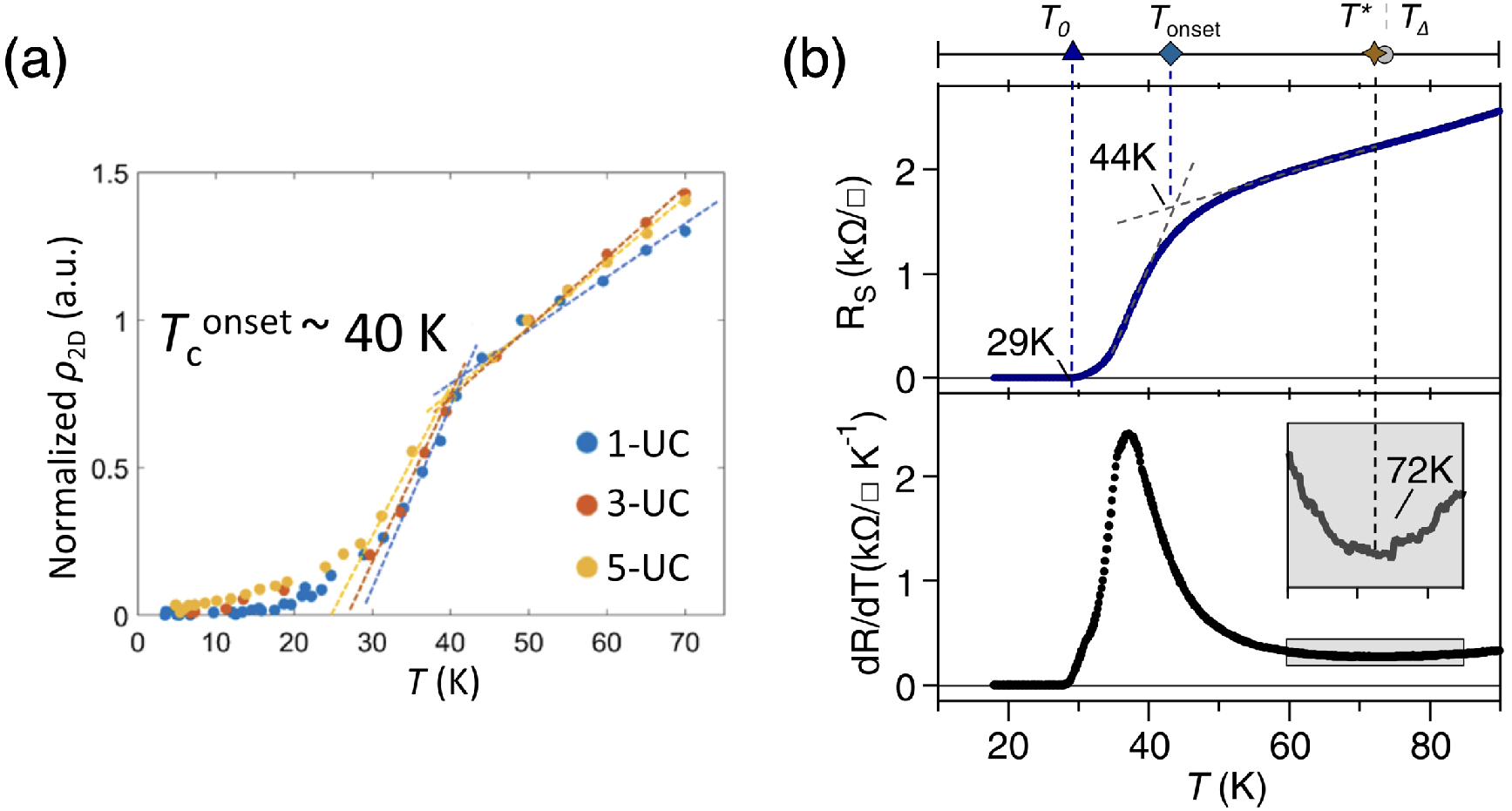}
\caption{Example of temperature dependence of sheet resistance of ultrathin FeSe/STO films. (a) Tokyo Institute of Technology group \cite{Pedersen}.
Dashed lines are guides for the eye.
(b) Cornell University group \cite{Faeth}   Here, the temperature dependence of d$R$/d$T$ is also shown.
$T_0$, $T_{onset}$, $T^*$ and $T_{\Delta}$ are zeero-resistance temperature, onset temperature of superconductivity, temperature where    d$R$/d$T$ becomes minimum, and gap closing temperature in the ARPES spectrum, respectively.
}
\label{newres}
\end{center}
\end{figure}

The most interesting superconductivity is the possible high-$T_c$ superconductivity of category 3.
Recently, there have been approaches to prepare ultrathin samples by exfoliating bulk single crystals, but the $T_c$ of these samples is very low\cite{BLei,Farrar,Xie,Zhu}.
Therefore, it is safe to assume that some interface effect as mentioned above is important. Figure~\ref{newres} shows an {\it in-situ} measured electrical resistivity data of ultrathin FeSe films on SrTiO$_3$(STO) substrates recently reported\cite{Pedersen,Faeth}.
The onset $T_c$ in electrical resistivity is 40-45 K, which is lower than the spectroscopic $T_c$ ($\simeq$ 65 K)\cite{QYWang,SHe}. The temperature at which the resistance becomes zero is much lower.
Some people believe that the low zero resistance $T_c$ of category 3 SC is natural from the BKT-transition point of view\cite{BKT,BKT2}, which is specific to two-dimensional SC\cite{Faeth}.
Others argue that the tailing of the temperature dependence of the electrical resistivity below $T_c$ is a characteristic of Bose metals, which have been much discussed recently\cite{YLi}.

\begin{figure}[htb]
\begin{center}
\includegraphics[scale=0.6]{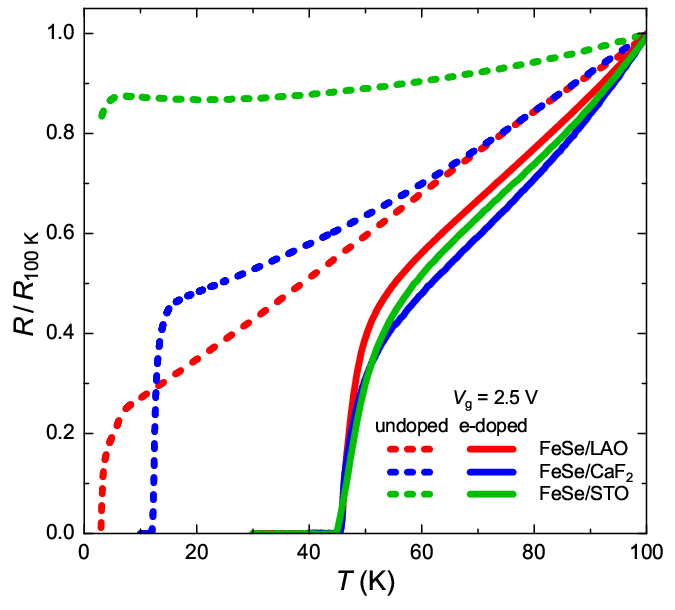}
\caption{Temperature dependence of electrical resistivity (normalized at 100 K) of FeSe thin films showing zero resistance at 46 K using electric double layer transistor (EDLT) \cite{Shikama2}.
The same zero resistance temperature is achieved on all substrates: LaAlO$_3$, SrTiO$_3$, and CaF$_2$.
}
\label{EDLT}
\end{center}
\end{figure}

On the other hand, we have recently achieved zero resistance of 46 K by doping electrons to a depth of about 10 nm from the surface using the electric field effect\cite{Shikama1,Shikama2} (Fig.~\ref{EDLT}). 
Therefore, we are very interested in the simple question of whether it is possible to achieve category 3 superconductivity in a more proper way, that is, with electrical resistance, and whether $T_c$ can be increased further.

\section{Studies using thin films}

\subsection{Merits of studies using films}
\begin{figure}[htb] 
\begin{center}
\includegraphics[scale=0.7]{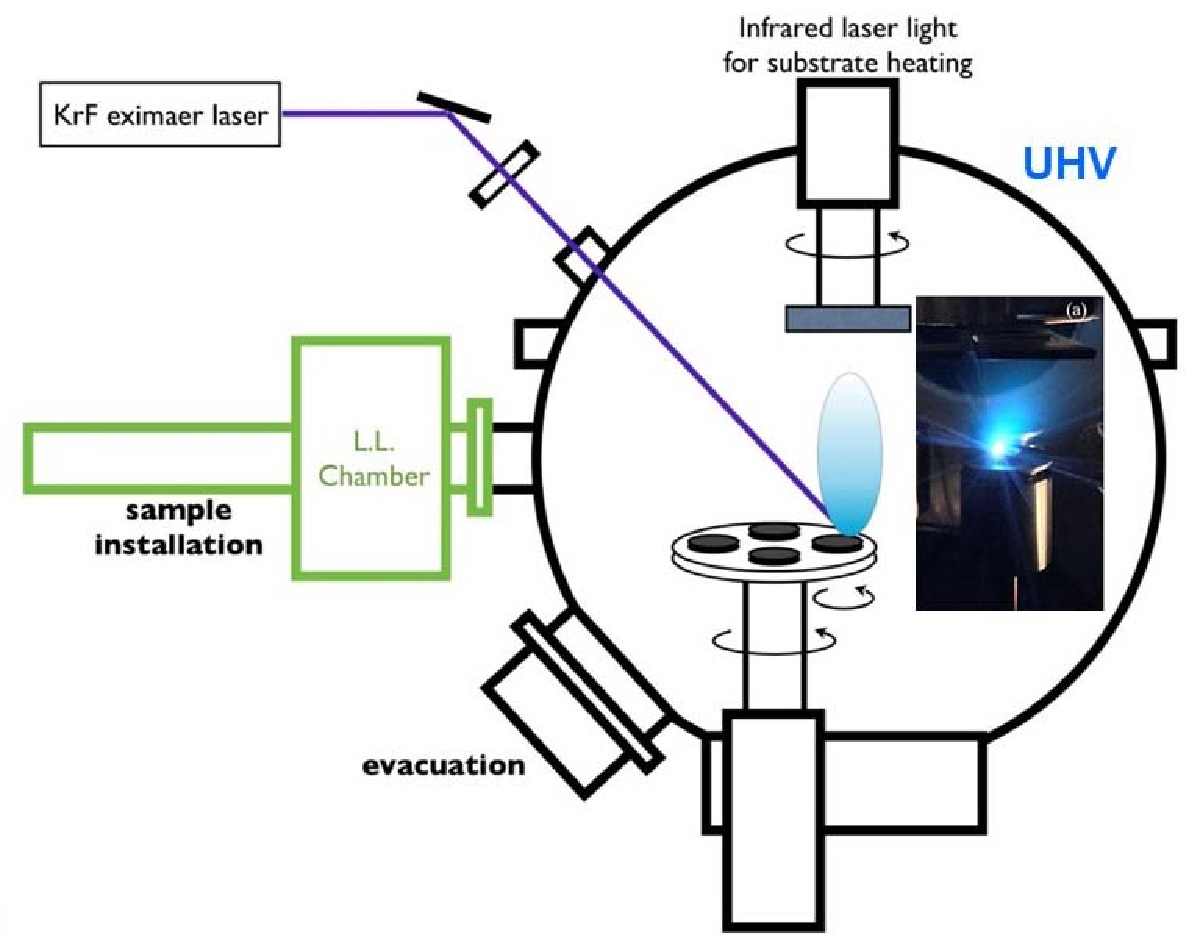}
\caption{Schematic figure of film formation by the PLD method.
}
\label{PLD}
\end{center}
\end{figure}

Thin film research has been widely conducted on iron-based superconductors from the early stages\cite{Hiramatsu,Haindl}.
Indeed, knowledge about thin film synthesis and processing is essentially important when developing devices using superconductors.
We have been working with this material since the relatively early stages of the discovery of iron-based superconductors, in the form of thin film fabrication by pulsed laser deposition (PLD) technique (Fig.~\ref{PLD}).

Thin-film research also has many advantages for fundamental research.
Namely, (1) large-area single crystals can be produced: this makes it easier to measure, for example, optical properties.
Specific examples will be introduced later.
(2) The non-equilibrium nature of the film fabrication process allows us to synthesize compositions that are difficult to synthesize in bulk: for example, we were the first to achieve Te substitution in the entire region $y$ with FeSe$_{1-y}$Te$_y$, which was previously difficult due to the phase separation, and achieved a $T_c$ 1.5 times higher than that of bulk in the conventional phase separation region\cite{Imai1,Imai2} (see also \cite{JZhuang}). We also achieved high-concentration S substitution in FeSe$_{1-x}$S$_x$, which led to the discovery of new anomalies\cite{Nabeshima1} (for these early results, see refs.\cite{ImaiR1,ImaiR2,ImaiR3} ).
(3) By changing the substrate for film fabrication, the strain can be systematically changed to a large extent to control the physical properties.
For example, in FeSe, by continuously changing the compressive strain from -1.5\% to +1.5\% tensile strain, it is possible to produce $T_c$ with a range from 12 K, which is 1.5 times the $T_c$ of the bulk at normal pressure, to 0 K (non-superconductor) \cite{Nabeshima2}.
Very recently, Huang {\it et al.} succeeded in obtaining $T_c=18.4$ K in FeSe filmss grown on LiF substrate, where very strong compressive strain is applied\cite{Huang2025}.
This result is in completely in line with the above mentioned results.
In particular, they proposed an empirical relation between $T_c$ and $c$ axis length of the film as $T_c\propto\sqrt{c-c_0}$ ($c_0=0.551$ nm)\cite{ZFeng2024}
(4) By making the film thinner, the effect of the interface can be seen, which has already been described.

\subsection{Same world vs another world}

When interacting with other researchers at academic conferences, {\it etc.}, we sometimes sense the following ``atmosphere".
``Thin films are another world, so we canot compare them with bulk single crystal data."
\begin{figure}[htb]
\begin{center}
\includegraphics[scale=0.5]{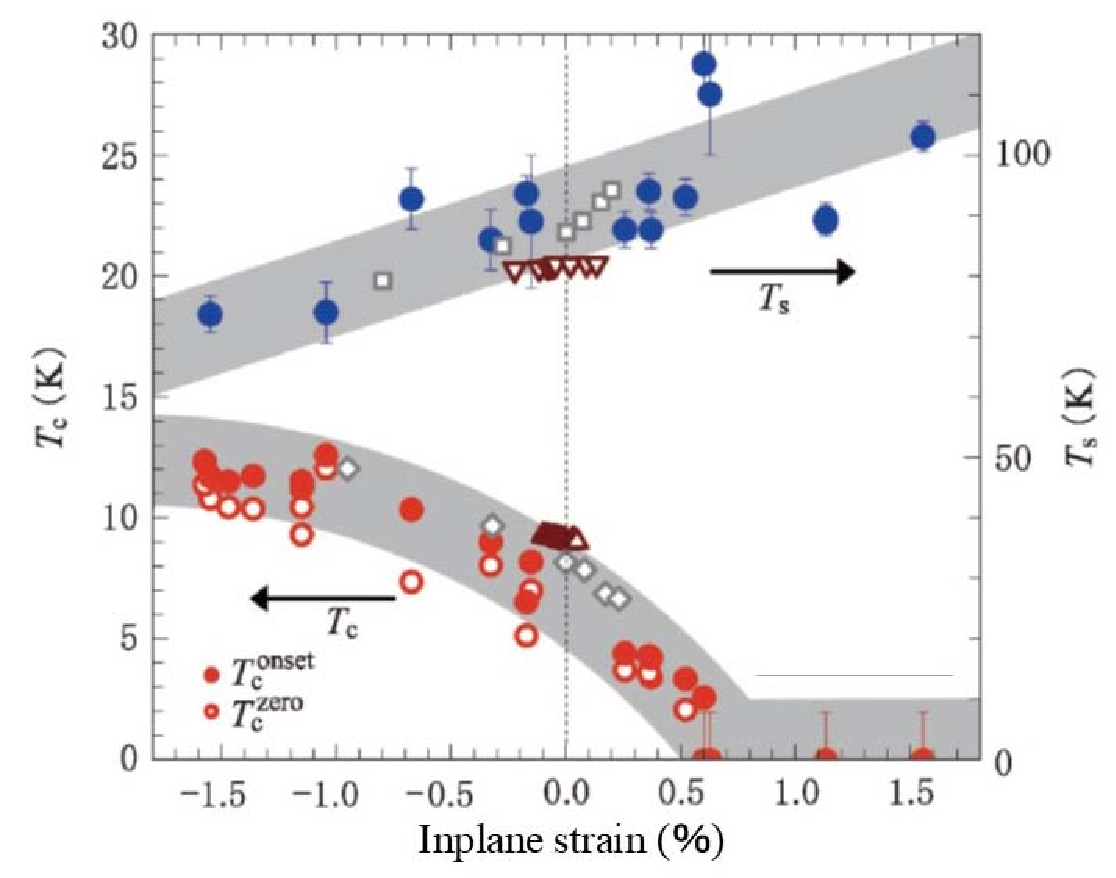}
\caption{Dependence of superconducting transition temperature $T_{\mathrm c}$ and nematic transition temperature $T_{\mathrm s}$ on in-plane strain. Illustrated are the results for uniaxial strain ($T_{\mathrm s}$ ($\bigtriangledown$) and $T_{\mathrm c}$ $(\Delta$))\cite{Ghini}, biaxial strain ($T_{\mathrm s}$($\opensquare$) and $T_{\mathrm c}$($\diamond$)) in bulk single crystals\cite{Nakajima1}, together with epitaxial thin film ($T_{\mathrm s}$(blue $\bullet$), $T_c^{onset}$ (red $\bullet$) and $T_c^{zero}$ (red $\circ$)) \cite{Nabeshima2}.
Hatched areas are guides for the eye.
}
\label{strain}
\end{center}
\end{figure}
I will show that this is a big misunderstanding by using concrete examples. 
To do this, let us take another look at the effect of strain, which I mentioned slightly above. 

Figure~\ref{strain} shows the results of a study on bulk single crystals of FeSe, in which uniaxial strain was applied to the sample to examine how $T_c$ and the nematic transition temperature $T_s$ change\cite{Ghini}, along with the results for our thin film samples\cite{Nabeshima2}.
When the strain changes from compression to tension, $T_c$ decreases, while the nematic transition temperature $T_s$ increases.
In this experiment, the strain was changed between +0.1\% and -0.2\%.
The changes in $T_c$ and $T_s$ of the thin film sample almost overlap with those of the bulk single crystal, but the range of strain changes in films is much wider than in bulk crystals, from +1.5\% to -1.5\%.
The results of an experiment in which biaxial strain was applied to a bulk single crystal are also shown in the same figure \cite{Nakajima1}.
A wider range of strain even in bulk crystals was successfully applied than in the results of uniaxial strain. 
Furthermore, the results of $T_c$ and $T_s$ are quantitatively consistent with those of the thin film.
Hall effect measurements have shown that the carrier density increases as the strain becomes compressive [59].
ARPES measurements comparing compressible strained, unstrained, and tensile strained samples also have confirmed that the overlap between the electron and hole bands increases as the strain changes from tensile to compressive, resulting in an increase in the number of carriers\cite{Phan}.
Therefore, the increase in $T_c$ shows a positive correlation with the increase in carrier density.
With these examples, it is clear that, when the parameter of strain is taken into account, epitaxial thin film samples and bulk single crystals are looking at the same world.
This will also be an important viewpoint when we look at electronic phase diagrams later.

Some data in films were once considered problems.
One example is the sign of the low-temperature Hall coefficient,
which is different between bulk and thin-film samples.
However, later research using thin-film samples prepared by exfoliating bulk single crystals\cite{Zhu} clarified that the Hall coefficient changes systematically with the film thickness, and that the two are continuously connected. In other words, it was shown that thin-film samples on substrates are not special.
It has been pointed out that the change in the Hall coefficient may be due to a change in the nematic domain size with thickness. Whatever the cause, as shown above, the superconducting transition temperature is almost the same between the bulk and the thin film, and the difference in the sign of the Hall coefficient at low temperatures is thought to have no relation to the SC of this system.

Keeping these in mind, in this review article, we will introduce some of the results obtained from various collaborative studies using our epitaxial thin film samples, which will be discussed in a comparative manner with related results in films and bulk crystals by others.
Furthermore, we will introduce characteristics unique to thin film samples.

\section{Category 1}

\subsection{Electronic phase diagram}

\begin{figure}[htb]
\begin{center}
\includegraphics[scale=0.7]{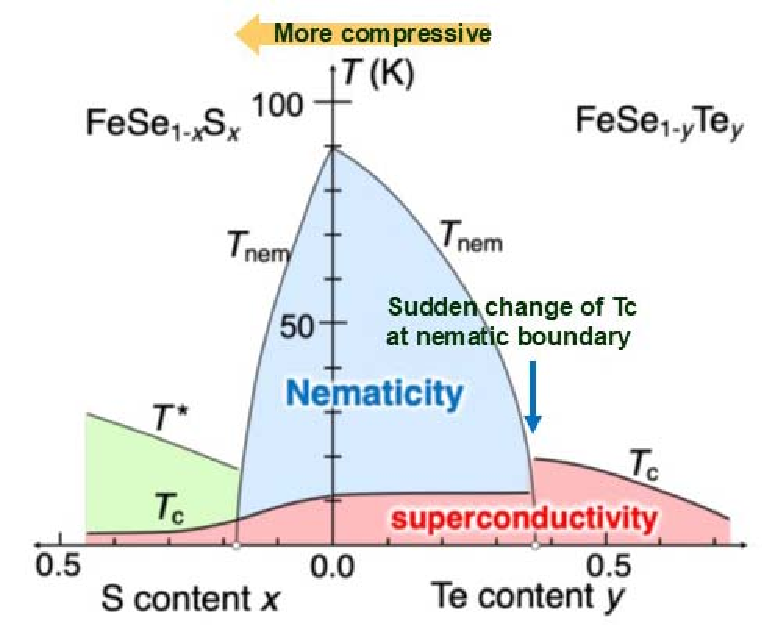}
\caption{Schematic phase diagrams of FeSe$_{1-x}$S$_x$ and FeSe$_{1-y}$Te$_y$ obtained in epitaxial thin film samples for category 1 SC.
$T_{nem}$, $T_c$ and $T^*$ represent nematic transition temperature, superconducting transition temperature and temperature where the temperature dependence of resistivity shows a kink, respectively.
}
\label{phasesche}
\end{center}
\end{figure}

Figure~\ref{phasesche} shows a schematic phase diagrams obtained so far for FeSe$_{1-y}$Te$_y$ and FeSe$_{1-x}$S$_x$ thin films\cite{Imai1,Imai2,Nabeshima1}.
This homologous element substitution has been attempted by many groups since the early days\cite{Hsu,Fang,Mizuguchi2,Watson1}. The substitution of S, which has a smaller atomic radius than Se, corresponds to the application of positive chemical pressure, while the substitution of Te, which has a larger atomic radius, corresponds to the application of negative chemical pressure.
Starting with FeSe, the nematic transition disappears regardless of whether the chemical pressure is positive or negative, but the behavior of $T_c$ at the nematic endpoint is contrasting.
For Te substitution, $T_c$ rises sharply to about 1.5 times the bulk $T_c$ at ambient pressure, while for S substitution, $T_c$ only decreases gradually with increasing S content and no significant change in $T_c$ is seen before and after the disappearance of the nematic transition.
\footnote{In S-sbstitute bulk single crystals, a small sudden decrease in $T_c$ was observed at the nematic boundary, and definite changes in superconductivity are also reported\cite{bulkSchange1,bulkSchange2}.}
These show that the role of nematic order in superconductivity is complicated.
In addition, a new anomaly is observed with high S
substitution, which appears as an upturn in the temperature dependence of electrical resistivity at $T^*$. 
As will be described in detail later, it turns out that this anomaly is due to the formation of short-range magnetic order.\footnote{No magnetic ordering has been reported in the bulk samples, but an upturn in the electrical resistivity has been reported in highly S$-$substituted samples, similar to that observed in thin film samples\cite{Shi_FeSeS}.} 
Therefore, it is necessary to investigate how the electronic state and carrier dynamics change within this phase diagram, especially with Te
substitution, which shows a steep change in $T_c$.
\begin{figure}[htb]
\begin{center}
\includegraphics[scale=0.7]{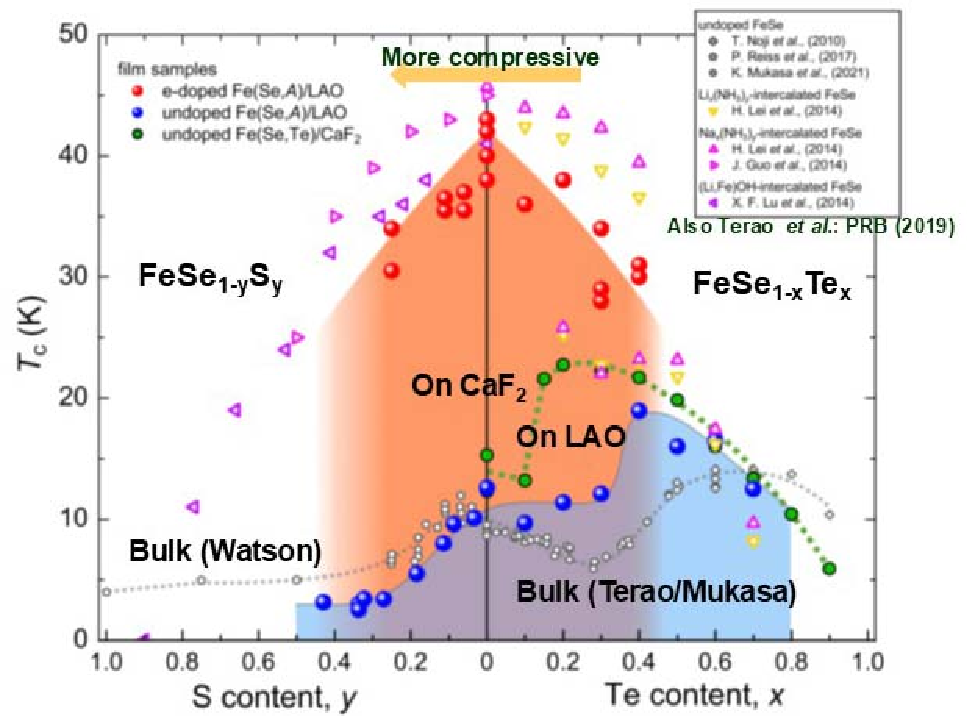}
\caption{Phase diagrams of FeSe${}_{1-y}$S$_y$ and FeSe$_{1-x}$Te$_x$ in thin film samples, including both category 1 and category 2 (adapted from ref.~\cite{Shikama2}). 
Blue balls and green closed circles are $T_c$s of category 1 films on LAO and CaF$_2$, respectively, and red balls are $T_c$s of ctegory 2 EDLT films. 
Violet colored and yellow colored open triangles are $T_c$s of category 2 samples taken from literatures descrfibed in the upper right corner. 
Small hollow black circles represent bulk data combined from several literature sources.
For sources of non-thin film data for bulk crystals and category 2, see the corresponding figures in the original papers referred in the text. 
Hatchings and dotted curves are only guides for the eye.
}
\label{phaseex}
\end{center}
\end{figure}

Recently, it has become possible to synthesize bulk single crystals in compositions that were previously considered to be in the phase separation region (Te substitution range of 10 \%$\sim$40\%)\cite{Terao,SLiu,Mukasa}.
Thus, it becomes possible to comparatively discuss the electronic phase diagrams between in films and bulk crystals also for Te substituted region.
In Fig.~\ref{phaseex} the $T_c$ data of bulk single crystals, thin films on LaAlO$_3$ (LAO), and thin films on CaF$_2$ are plotted in the same figure\cite{Shikama2}. (In this figure, also the $T_c$ data of category 2 are shown, which will be described later.). 
As emphasized at the end of the previous section, when the parameter of strain is kept in mind, the data for bulk single crystals, thin films on LAO, and thin films on CaF$_2$ appear to change systematically\footnote{When a film is formed on LaAlO$_3$, tensile strain is usually applied, but in the sample shown in this figure, a small amount of compressive strain is applied\cite{Nabeshima1}.}.
In all three samples, an increase in $T_c$ was observed in the non-nematic region for Te substituted samples.
The broad dome shape of the $T_c$ {\it vs} $x$ in bulk crystals leads to an argument that  $T_c$ enhancement is caused by the nematic fluctuation around the nematic quantum critical point (QCP)\cite{bulkQCP}.
However, particularly noteworthy is that in the thin film samples, the increase in $T_c$ after the disappearance of the nematic transition is abrupt when viewed as a function of Te content, $x$.
Below we will show that this is due to the big change in the electronic structure at the nematic end point and that the enhancement of $T_c$ by the nematic fluctuation, if any, is very small.

Recently, a group led by Professor Kui Jin of IOP-CAS reported that they introduced a combinatorial approach to the PLD method and succeeded in synthesizing a single phase over the entire Te composition range\cite{Qihong}. 
The phase diagram obtained shows a single dome shape with a maximum $T_c$ in the middle as the Te content, $x$ increases from zero to 1, and no anomalies are observed at the nematic boundary. 
They also observed the nematic transition in the phase diagram almost the same as was reported in ref.~\cite{Imai1}.
Another characteristic in their data is that the Fe deficiency of 20\% is common across all regions.
From these findings, it may be possible that the large change in $T_c$ before and after the nematic boundary reported above is due to a significant change in the amount of Fe before and after the boundary.
Although the combinatorial method is suitable in synthesizing a wide range of samples simultaneously, it is believed that the sample preparation conditions have not been optimized for each composition, and the resultant amount of iron deficiency is as high as 20\%.
Also, if the 1.5-fold change in $T_c$ before and after the nematic boundary is due to a change in the amount of Fe, it will necessarily be reflected in the change in the lattice constant.
Therefore, it seems unreasonable to attribute the large change in $T_c$ at the nematic boundary to a large change in the amount of Fe vacancy.
Although the origin of the different phase diagram is not clear even now, the resistivity measurement in combinatorially fabricated samples may be the average of areas with different Te composition $x$ even for a very narrow range.
This can obscure possible sharp changes in the $T_c$ {\it vs} Te concentration relation.

Sato {\it et al} realized FeSe$_{1-x}$Te$_x$ films for whole $x$ values between 0 and 1 on a magnetic topological insulator CdTe substrate (grown by MBE)\cite{Sato2024}.
All samples including FeTe show superconductivity, and the phase diagram is rather different from the above-mentioned studies, suggesting rather different physics might be involved in their data.
What is interesting is the lattice constants of their films obey the Vegard' s law between bulk FeSe and FeTe, which means no stress/strain presents in all films.
A succeedig study on the same substrate material\cite{Sato2025a} are published within a story of quantum criticality.
Superconductivity of FeTe was also reported on monolayer FeTe on another topological insulator (Bi,Sb)$_2$Te$_3$ substrate (also grown by MBE)\cite{Yi2024}. 
Realization of superconductivity also for FeTe might be related to higher-order epitaxy partly\cite{Sato2025b}.

\subsection{Pure nematicity without any lattice displacement}
\begin{figure}[htb]
\begin{center}
\includegraphics[scale=0.4]{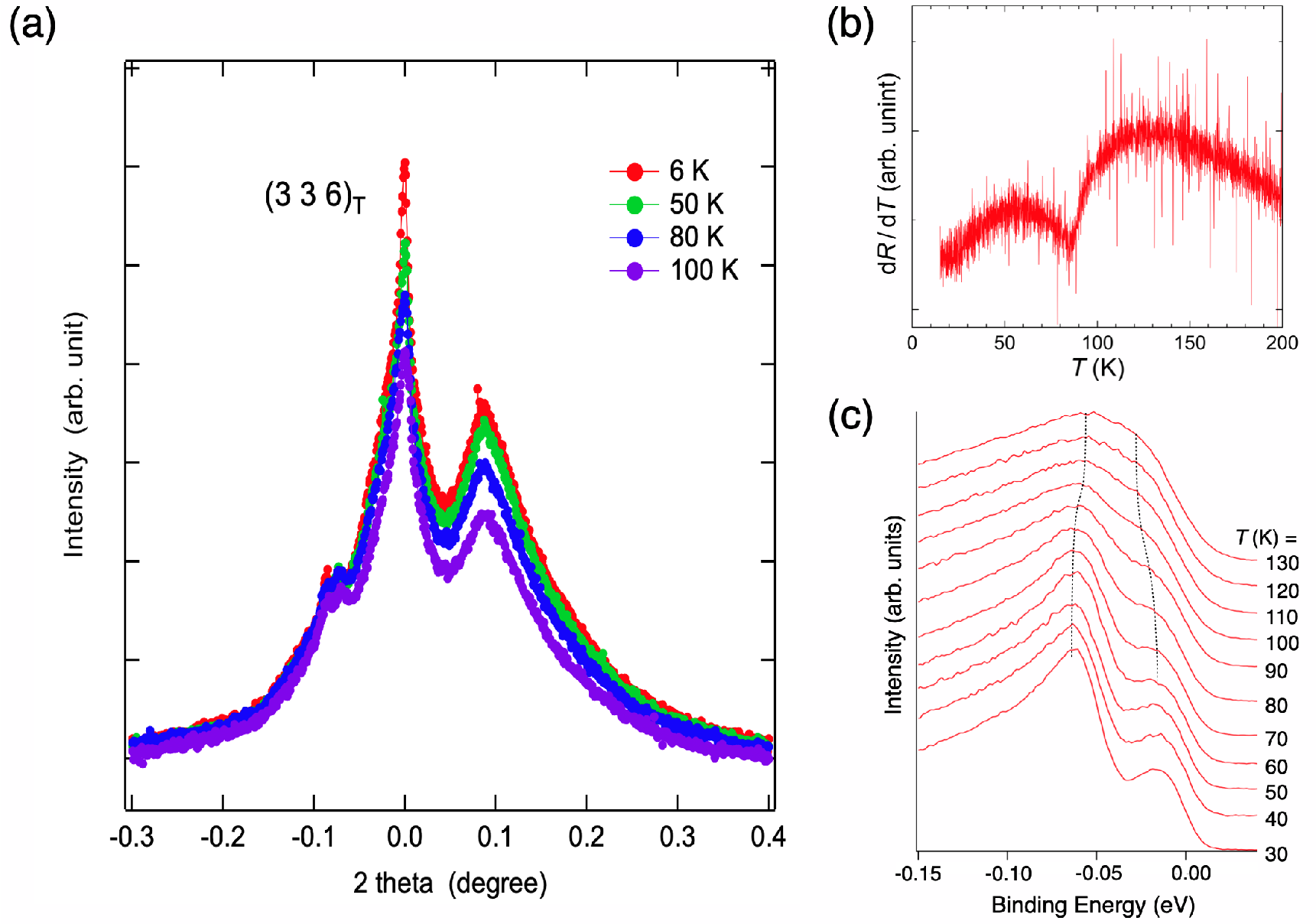}
\caption{(a) Temperature dependence of (336) reflection, (b)
temperature dependence of electrical resistivity differentiated with respect to temperature, and (c)
temperature dependence of ARPES spectra for a thin FeSe film (thickness 50 nm) on LAO (Ref.~\cite{Kubota}. )
}
\label{pure}
\end{center}
\end{figure}

Figure~\ref{pure}~(a) shows the temperature dependence of the (336) reflection of FeSe deposited on LAO\cite{Kubota}. 
The Miller indices are for the tetragonal unit cell at the higher temperature.
In bulk single crystals, FeSe undergoes a structural phase transition at about 90 K\cite{ortho}.
In contrast, in Fig.~\ref{pure}~(a), no change in the diffraction peaks is observed even when the temperature is changed across 90 K. The multiple peak structure in Fig.~\ref{pure}~(a) is due to cracks in the substrate, and a usual $\theta - 2\theta$ scan confirms that there is only one peak even at the lowest temperature measured, 6 K.
Therefore, it is clear that the lattice does not show any phase transition in this sample.
On the other hand, as shown in Fig.~\ref{pure}~(b), the temperature dependence of the electrical resistivity clearly shows an anomaly corresponding to the transition at 90 K, and ARPES measurements show that the bands corresponding to the $d_{yz}$ orbital and $d_{xz}$ orbital clearly change with temperature above and below 90 K (Fig.~\ref{pure}~(c)). Even above 90 K, the energies of the two bands do not become the same, but remain different, which is understood to be due to spin-orbit interactions.
It is surprising that, although the electronic state (band structure) clearly shows a transition, the lattice does not show any change at all.

In many iron-based superconductors, the non-equivalence of electron orbitals accompanying such a structural phase transition is thought to be electron-driven, since the energy splitting of the electron system is much larger than that expected from the magnitude of the observed lattice displacement.
Therefore, this transition is called the nematic transition and is considered one of the important keywords in discussing the physics of iron-based superconductors\cite{nematic}.
In that sense, the non-equivalence of electronic states without any lattice displacement observed here is consistent with that trend. However, it is still surprising that a nematic transition has been observed in such a pure form.

The effective interaction between electrons and lattices, $\lambda_{eff}$, is given by the frequency of the phonon, $\omega_{ph}$, and the bare electron-lattice interaction. $\lambda$ as
\begin{equation}
\lambda_{eff}\sim\frac{\lambda^2}{\omega_{ph}}
\end{equation}
Therefore, if we assume that the lattice deformation is hindered by the influence of the substrate in the thin film and the resultant phonon frequency is increased, $\lambda_{eff}$ will become small, and the observed pure nematic transition can be qualitatively understood.
In {\it sec. 2.2}, we wrote that ``thin films and bulk materials see the same world" and ``if strain is kept in mind, the electronic phase diagrams of thin films and bulk materials can be viewed continuously."
However, this is not the case when it comes to the coupling between the electron system and the lattice system, and it is reasonable to think that in thin film samples, a purer form of the electronic system that is not affected by interactions with the lattice is seen.
We believe that this is related to the fact that in the electronic phase diagram, the change in $T_c$ appears to be more abrupt in the thin film samples than in bulk samples when the Te content, $x$, is changed to suppress the nematic transition.

\subsection{Normal state}

\subsubsection{Changes in electronic state observed by ARPES}

\begin{figure}[htb]
\begin{center}
\includegraphics[scale=0.55]{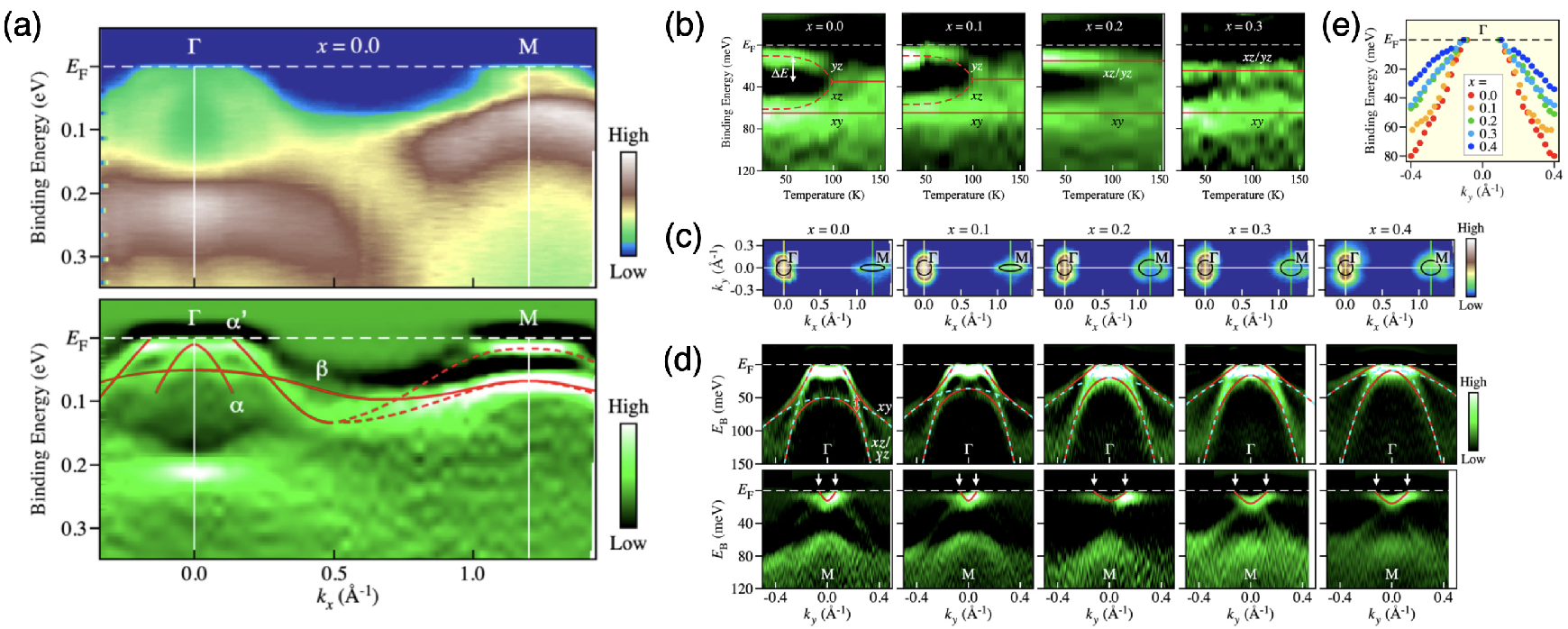}
\caption{
ARPES measurements of FeSe$_{1-x}$Te$_x$ thin film samples on CaF$_2$ substrates. (a) the APRES intensity (upper panel)
and the corresponding second-derivative intensity (lower panel) of an $x=0$ (FeSe) sample,
measured along the $\Gamma -M$ line at $T =$30 K with linearly polarized 21.2-eV photons. 
(b) The temperature dependence of the second-derivative ARPES intensity at the M point for samples with various Te contents $x$.
The red curves are guides for the eye to trace the band dispersions. 
(c) Te-concentration dependence of the ARPES-intensity map at $E_F$ as a function of the two-dimensional wave vector for
FeSe$_{1-x}$Te$_x$/CaF$_2$, with $x =$ 0.0,  0.1, 0.2,  0.3, and 0.4, at $T =$ 30 K. 
Black curves are the schematic of FSs.
(d) The near-$E_F$ second-derivative ARPES intensity obtained at $T =$ 30 K along yellow and green lines shown up in (c), 
Cyan and red curves are guides for the eye to trace the band dispersions before and after hybridization, respectively.
The band dispersions before the hybridization were estimated by extrapolating the dispersions determined in the
($E$, $\bi{k}$) region away from the intersection point.
(e) Comparison of the experimental near-$E_F$ band dispersions
extracted by tracing the peak position of ARPES spectra. (\cite{Nakayama1})  
}
\label{ARPES}
\end{center}
\end{figure}

\

Let us see how the electronic state changes with Te substitution.
Figure~\ref{ARPES} shows ARPES results for a series of FeSe$_{1-x}$Te$_x$ samples grown on CaF$_2$\cite{Nakayama1}.
Since fresh cleavage surfaces are required for ARPES experiments, we had difficulty cleaving thin film samples, and ended up trying more than 70 thin films to obtain this series of data.
The band dispersion of FeSe (Fig.~\ref{ARPES}~(a)) obtained from the second derivative of the spectrum is almost the same as that obtained for bulk samples\cite{Phan,bulk_ARPES}, and each band can be identified as shown in the figure.
As shown in Fig.~\ref{ARPES}~(b), in the samples with $x =$ 0 and 0.1, the $d_{yz}-$like band and the $d_{xz}-$like band split, but merge as the temperature rises. In other words, the nematic transition is clearly observed.
In contrast, in the samples with $x =$ 0.2 and 0.3, these bands barely change with temperature, and the nematic transition disappears.
Correspondingly, the shape of the Fermi surface at the M point at low temperatures is a long ellipse in the nematic state, but in the non-nematic samples, it changes to an almost circular shape (Fig.~\ref{ARPES}~(c)).
On the other hand, no such significant change is observed in the shape of the Fermi surface at the $\Gamma$ point.
However, looking at the band dispersion in Fig.~\ref{ARPES}~(d), it can be seen that the energy of the $d_{xy}-$like orbital increases with Te substitution, and hybridizes with the $d_{xz}$ orbital due to spin-orbit interaction, and the Fermi velocity $v_F$, {\it i.e.}, the density of states $N\simeq 1/v_F$, changes significantly before and after the nematic transition (Fig.~\ref{ARPES}~(e)). 
This is thought to be the direct consequence of the disappearance of the nematic transition and the resultant large increase in $T_c$ due to the substitution of Te.

\subsubsection{Carrier dynamics}

\begin{figure}[htb]
\begin{center}
\includegraphics[scale=0.8]{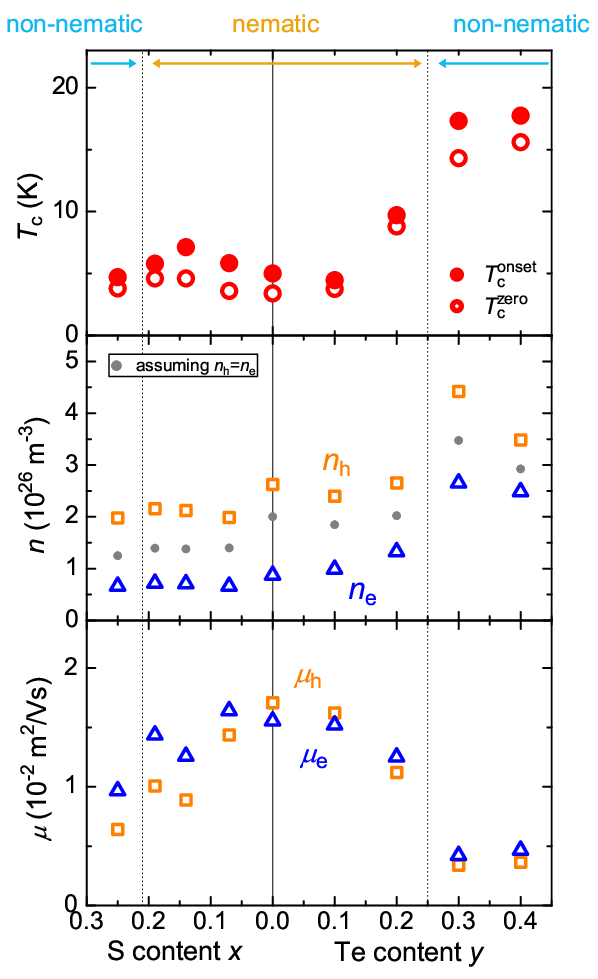}
\caption{Composition dependence of $T_c$ (top), as well as the electron density, $n_e$, hole density, $n_h$ (middle), and respective mobilities, $\mu_e$ and $\mu_h$ (bottom) obtained from Hall effect measurements in FeSe$_{1-x}$S$_x$ and FeSe$_{1-y}$Te$_y$ thin film samples\cite{Nabeshima3}.
Grey points in the middle panel are those obtained under the assumption, $n_h=n_e$.
The dotted lines show the nematic boundaries.
}
\label{transport}
\end{center}
\end{figure}

\

The increase in the number of carriers due to the disappearance of the nematic transition in Te substituted films is also reflected in the carrier dynamics.

First, let us look at the dc transport phenomenon. Figure~\ref{transport} shows the carrier density calculated from the Hall effect as a function of S and Te substitution\cite{Nabeshima3,Sawada}. \footnote{Due to the usage of figures in the original papers, the notations for the substituted contents, $x$ and $y$ are frequently interchanged between S and Te.  We ask readers for patience not to be confused.}
To evaluate these, a two-carrier model was used, in which holes and electrons are each represented by one carrier\footnote{In this material, two or more bands actually contribute to conduction. 
However, unless the phase between the bands matters, it is not a bad approximation to represent the hole band and the electron band by one band each.}.
In these samples, the Hall voltage is proportional to the magnetic field and there is no region of nonlinear dependence.
So, when evaluating the number of carriers and mobility, some assumptions must be introduced
(for example, the number of holes and the number of electrons are assumed to be equal), which should introduce the ambiguity.
We proceeded the analysis by referring to the carrier concentration obtained in our film samples by the complex conductivity measurement in THz region\cite{Yoshikawa}, where the frequency dependence of the longitudinal component of the conductivity tensor and the Hall component is very well expressed by the Drude type.
In this way, the carrier-density values obtained by fitting the frequency dependence of the conductivity tensor are highly reliable.

Returning to Fig.~\ref{transport}, when the nematic transition disappears in the Te substitution, the carrier density increases for both electrons and holes, whereas in the S substitution, the carrier density hardly changes even when the nematic transition is on either side.
In this way, the composition dependence of the increase or decrease in carrier density in Te substituted samples corresponds well to the composition dependence of the increase or decrease in $T_c$.

\begin{figure}[htb]
\begin{center}
\includegraphics[scale=0.6]{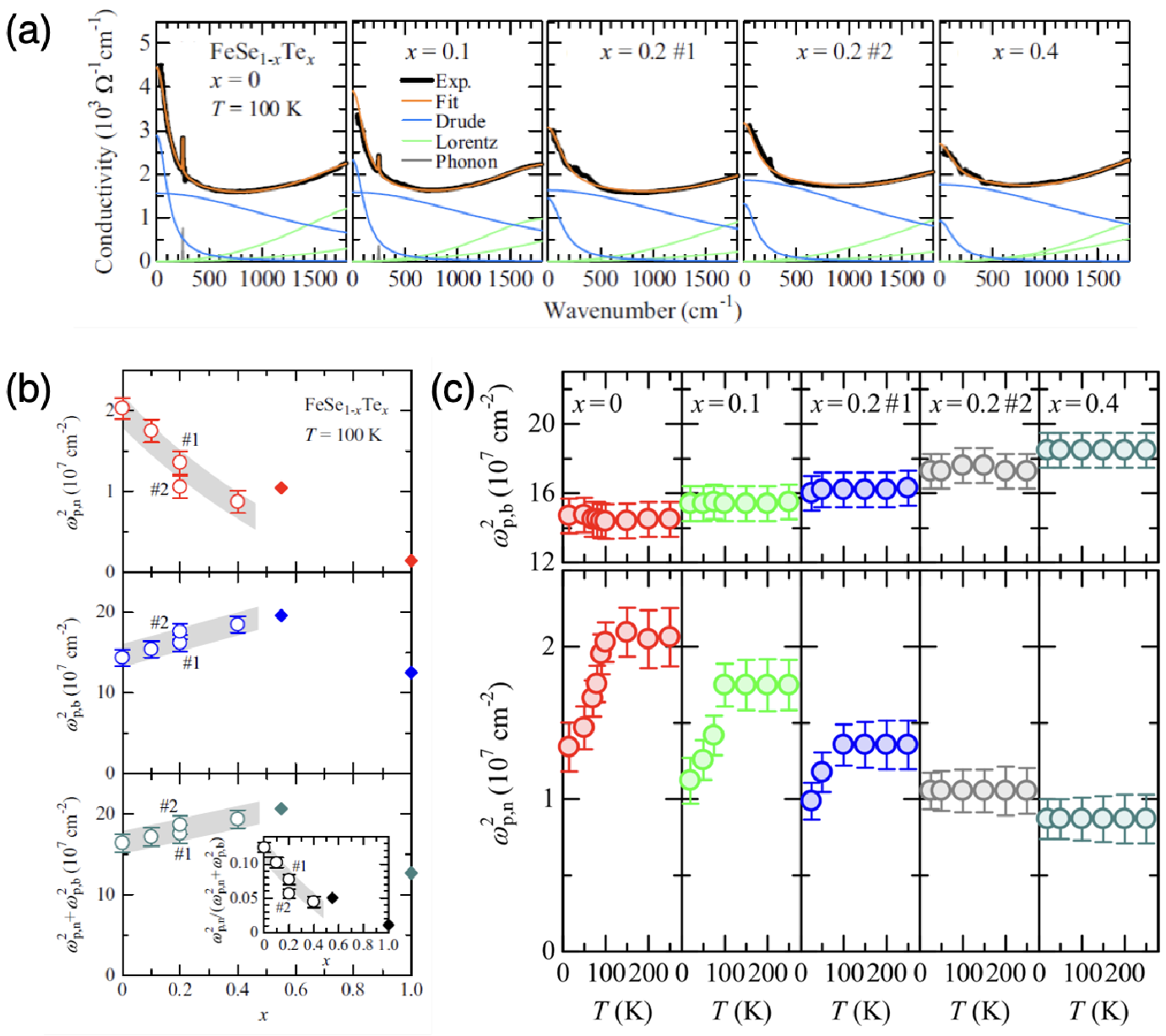}
\caption{ For FeSe$_{1-x}$Te$_x$ thin films on CaF$_2$: (a) Frequency dependence of electrical conductivity in the optical region for samples of various Te compositions. Data for each sample is expressed as the sum of the two Drude components, represented by cyan and green curves. (b) Results of the decomposition of the optical conductivity
spectra for FeSe$_{1-x}$Te$_x$ as a function of $x$ at T = 100 K. (Top) The weight of the narrow Drude component $\omega^2_{p,n}$ and (Middle) those of the broad Drude component $\omega^2_{p,b}$. (Bottom) Evolution of the total Drude weight $\omega^2_{p,n}+\omega^2_{p,b}$. The inset shows the composition dependence of the fraction of the narrow Drude weight $[\omega^2_{p,n}/(\omega^2_{p,n}+\omega^2_{p,b})]$. 
Gray bands are guides to the eye.
The data for x = 0.55 and 1.0 measured on single crystals are also plotted.
(c) Temperature dependence of $\omega^2_{p,n}$ and $\omega^2_{p,b}$ for FeSe$_{1-x}$Te$_x$
($x =$ 0, 0.1, 0.2 No. 1, 0.2 No. 2, and 0.4).
The reduction of $\omega^2_{p,n}$ at low temperatures indicates the presence of the nematic phase
transition.\cite{Nakajima3}. 
}
\label{optical}
\end{center}
\end{figure}

Next, we introduce the ac conductivity up to the optical frequencies in a series of Te-substituted samples\cite{Nakajima2,Nakajima3} (Fig.~\ref{optical}).
As shown in Fig.~\ref{optical}~(a), in all samples, we can see that the frequency dependence of the real part of the electrical conductivity can be expressed by two components: a narrow Drude component corresponding to coherent carriers and a broad Drude component corresponding to incoherent carriers.
This behavior is common to other iron-based superconductors\cite{FBSoptical1,FBSoptical2,FBSoptical3,FBSoptical4} and is considered to be a typical feature of correlated electron systems.
Figure~\ref{optical}~(b) shows the composition dependence of each oscillator strength (plasma frequency).
The broad component, $\omega^2_{p,b}$ increases with increasing Te content, while the coherent component, $\omega^2_{p,n}$ decreases.
Therefore, it can be said that Te substitution strengthens the electron correlation. 
In addition, as shown in Fig.~\ref{optical}~(c), in samples that show a nematic transition, the coherent component $\omega^2_{p,n}$ decreases rapidly below the nematic transition temperature, indicating that superconductivity and nematic order conflict with each other.
This also coincides with the ARPES results.

\subsubsection{Band calculation}

\begin{figure}[htb]
\begin{center}
\includegraphics[bb=0 0 676 568,scale=0.5]{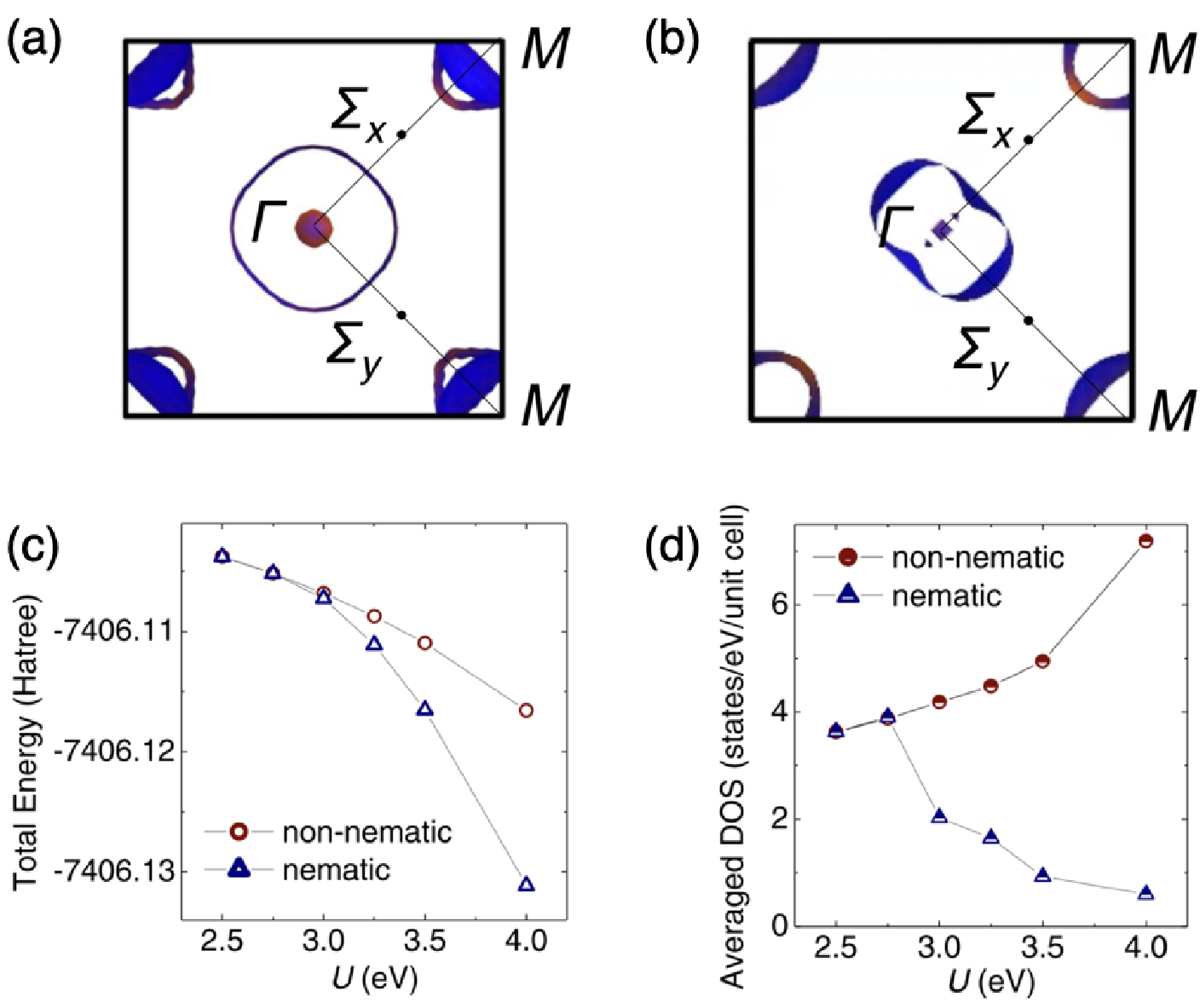}
\caption{ Fermi surfaces and energy difference of the nematic and non-nematic states of iron chalcogenides obtained by DFT calculations. (a) Fermi surface of the non-nematic state, (b) Fermi surface of the nematic state, (c) Ground-state energy of FeSe as a function of $U$ in the nonnematic and nematic phases. Soiid curves are guides for the eye. (d) Averaged density of states per unit cell of
FeSe as a function of $U$ in the nonnematic and nematic phases. Soiid curves are guides for the eye.\cite{Kurokawa}. 
}
\label{DFT}
\end{center}
\end{figure}

\

DFT calculations were performed to qualitatively understand these experiments\cite{Kurokawa}. 
Many DFT calculations have been performed on the FeCh system since the early days.
However, all of them have a common problem of producing Fermi surfaces that are much larger than those observed in experiments, and many attempts have been made to improve them, including the inclusion of electron correlation\cite{XLiu,Coldea,Kreisel}.
Among them, the calculations by the Tsinghua University group were epoch-making\cite{XLong}.
In all previous calculations, it was necessary to assume some kind of magnetic order to obtain nematic order,
whereas this calculation succeeded in obtaining a ground state with nematic order without assuming any magnetic order.
This calculation shows that as the electron correlation $U$ increases, the nematic state becomes increasingly stable. 
We calculated the Fermi surface and energy using the same method, and confirmed that their results were reproduced (Fig.~\ref{DFT}~(c)).
We then calculated the density of states, which showed that the difference in DOS between the two states increases as $U$ increases (Fig.~\ref{DFT}~(d)).
As mentioned above, the optical measurements show that Te substitution strengthens the electron correlation, whereas the direction of S substitution can be considered to weaken the electron correlation.
Thus, it is natural that the nematic transition is suppressed by S substitution.
On the other hand, Te substitution leads to an increasingly stable nematic transition, and the suppression of the nematic transition cannot be understood within this framework. However, once the nematic order is suppressed in the large $U$ regime, a large increase in DOS is expected. This is thought to qualitatively explain the rapid increase in $T_c$ due to the suppression of the nematic transition.

The results in Fig.~\ref{DFT} shows what takes place when only $U$ is changed.
So to understand the suppression of the nematic transition by Te substitution, we will need to find a solution in either the change in the band structure or Fermi surface itself due to the Te substitution, the change in magnetic fluctuations, or the change in spin-orbit interaction, {\it etc}.

Also inspired by the calculations of Lang {\it et al.,}, Yamada-Toyama performed a multipole analysis\cite{Yamada} and concluded that the emergence of a nematic ground state from a nonmagnetic state is due to the interaction between an antiferromagnetic hexapole state in the E representation and a multipole state in the B2 representation in the presence of Coulomb repulsion $U$.

\subsubsection{Magnetic property}

\begin{figure}[htb]
\begin{center}
\includegraphics[bb=4 0 593 679,scale=0.5]{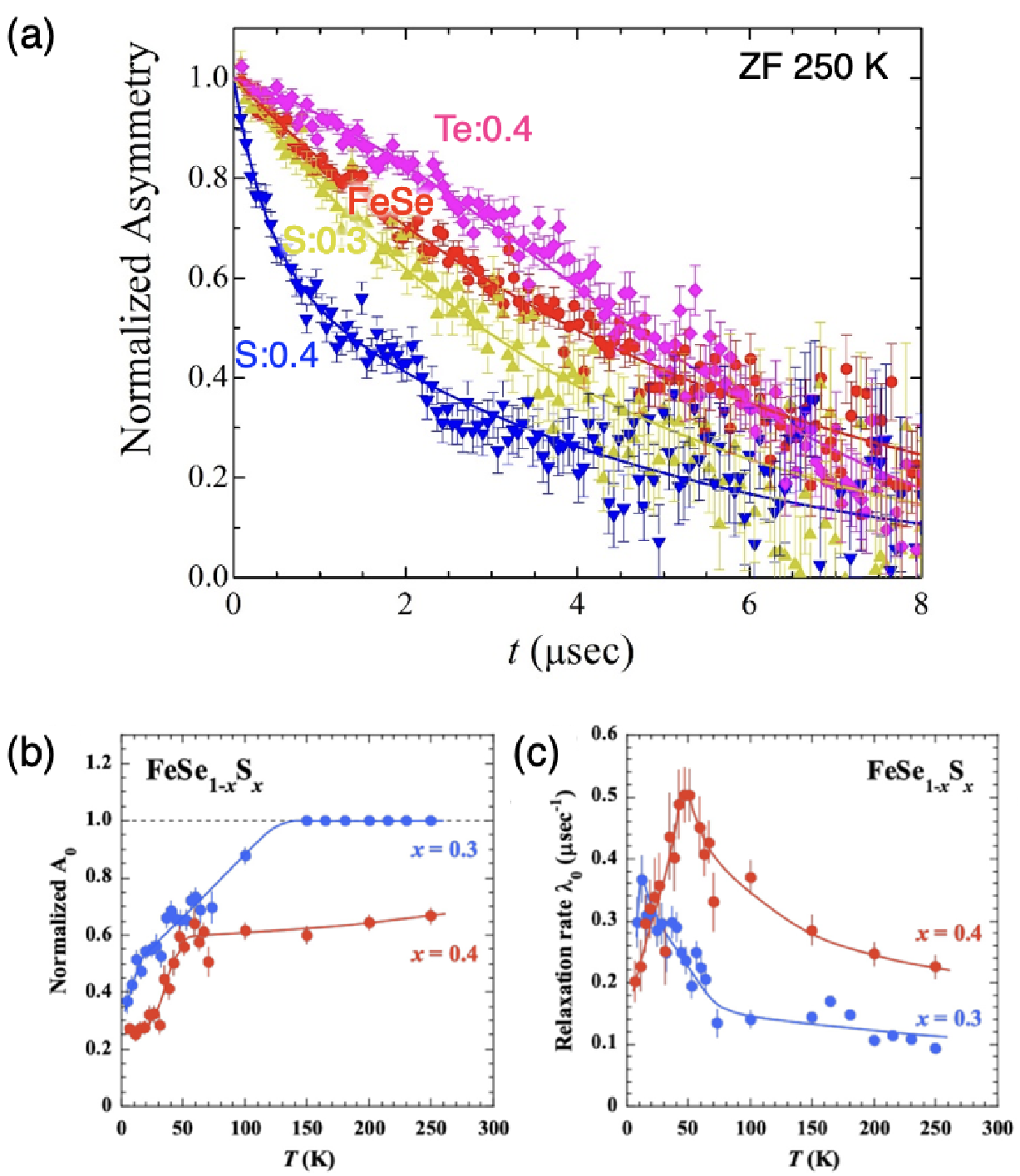}
\caption{$\mu$SR results for FeSe, FeSe$_{0.6}$S$_{0.4}$, and FeSe$_{0.6}$Te$_{0.4}$. (a) Asymmetry relaxation for FeSe, FeSe$_{0.6}$S$_{0.4}$, and FeSe$_{0.6}$Te$_{0.4}$, measured in zero-magnetic-field condition at 250 K. (b) Temperature dependence of the initial asymmetry, $A_0$ obtained at 50 G for FeSe$_{1-x}$S$_x$ films with $x =$ 0.3 and 0.4. Solid curves are guides for the eye.. (c) Tem;erature dependence of the relaxation rate of muon spins of the slow component, $\lambda_0$ obtained at 50 G for FeSe$_{1-x}$S$_x$ films with $x =$ 0.3 and 0.4. Solid curves are guides for the eye. ((b) and (c) are from \cite{Nabeshima4}.)
}
\label{muSR}
\end{center}
\end{figure}

\

In iron-based superconductors, magnetic fluctuation is an important keyword as well as nematicity. 
When trying to investigate the magnetic properties of thin-film samples, neutron scattering and NMR cannot be used due to insufficient signal strength.
The only possible method at present is $\mu$SR. In particular, the only place where thin-film $\mu$SR is possible is the Paul Scherrer Institute (PSI) in Switzerland, which can implant low-energy muons.
In $\mu$SR experiments, information about the magnetic state inside the sample is obtained by fitting the relaxation of the asymmetry $A$ of forward-scattered and backward-scattered muons as a function of time.

Figure~\ref{muSR}~(a) shows the time dependence of asymmetry for FeSe, FeSe$_{0.6}$S$_{0.4}$, and FeSe$_{0.6}$Te$_{0.4}$\cite{Adachinew,Nabeshima4}. 
The asymmetry decreases exponentially in the FeSe and S-substituted samples, which indicates the presence of short-range magnetic fluctuations in the samples.
The existence of antiferromagnetic fluctuations in FeSe has been shown by NMR experiments on bulk single crystals \cite{Takigawa}, which is in line with this $\mu$SR data.
The faster relaxation in the S-substitute sample suggests that the magnetic fluctuations are stronger. 
On the other hand, in the Te-substituted sample, the relaxation curve is convex upward, which is expressed by a Gaussian.
This is typical of a weak magnetic fluctuation, and we have to think that although the Te substitution strengthens the electron correlation, the magnetic fluctuations are weakened.

Figure~\ref{muSR}~(b)(c) shows the temperature dependence of the asymmetry amplitude and relaxation rate for the samples with S: 0.3 and 0.4 \cite{Nabeshima4}. 
For example, in the sample with S: 0.4, the asymmetry decreases rapidly at 50 K when the temperature decreases. 
On the other hand, the relaxation rate has a peak at the same temperature and decreases rapidly at lower temperatures.
These results indicate that short-range magnetic order is formed in this sample at temperatures below 50 K. 
The same phenomenon has also been observed in the sample with S:0.3. 
Remembering that an upturn in the electrical resistance was observed in these samples at this temperature, this corresponds to the occurrence of short-range magnetic order. 

We are continuing the investigation of $\mu-$SR properties of FeSe$_{1-x}$S$_{x}$ and FeSe$_{1-y}$Te$_{y}$ systematically.  It is turning out that the magnetic properties of FeCh epitaxial thin films are generally very complicated, which will be submitted soon\cite{Adachinew}.

\subsubsection{Summary of properties in the normal state}

\

To summarize the normal state of the samples in category 1,

(1) A pure nematic transition without any lattice distortion is observed in thin film samples, making them rare examples in which the pure response of the electron system free from interaction with the lattice can be observed.

(2) Various experiments have confirmed that both electrons and holes are involved in dynamics, and the two-carrier model works for quantitative description.

(3) Te substitution significantly changes the electronic state near the Fermi level. In the nematic state, only the $d_{yz}$ and $d_{xz}$ orbitals are near the Fermi level, but Te substitution increases the energy of the $d_{xy}$ orbitals, which then hybridize with the electronic states that make up the Fermi surface. 
In other words, this is orbital switching due to Te substitution. In relation to this, the density of states (and carrier density) at the Fermi level changes significantly. 

(4) Te substitution increases the electron correlation, but does not strengthen the magnetic fluctuations.

(5) All experiments, including ARPES, optical conductivity, dc magneto-transport study, and strain effect, show that superconductivity and nematic order are inversely related.

(6) In FeSe and S-substituted samples, short-range magnetic fluctuations exist, but in high-concentration S-substituted samples, they become ordered at low temperatures (short-range magnetic order).

(7) The $T_c$ of the superconductivity is positively correlated with the carrier density/ density of sttes at $E_F$ in the normal state.
This is probably the direct origin of the abrupt increae in $T_c$ when Te content $x$ crosses the nematic boundary. 

\subsection{Superconducting state}

The most important information in the superconducting state is the structure (symmetry) of the superconducting gap.
There are reports in bulk single crystals for certain specific compositions.
It is highly anisotropic in FeSe\cite{Sprau,CLSong}, completely nodeless in 50\% Te-substituted samples\cite{Hanaguri1}, and the gap structure changes with the disappearance of the nematic transition in S-substituted samples\cite{bulkSchange2,Hanaguri2}. Naturally, there is interest in how the gap changes around the endpoint of the nematic transition with Te substitution.
For this reason, we have attempted to measure complex conductivity in our film samples.
However, there is no established method for measuring complex conductivity in superconducting thin-film samples that are thinner than the magnetic penetration depth.
Before discussing the superconducting state of the FeCh system, we will briefly introduce the current status of complex conductivity measurements in superconducting thin-film samples.

\subsubsection{Complex conductivity measurement of thin-film samples
 }

\begin{figure}[htb]
\begin{center}
\includegraphics[bb=5 0 842 570,scale=0.45]{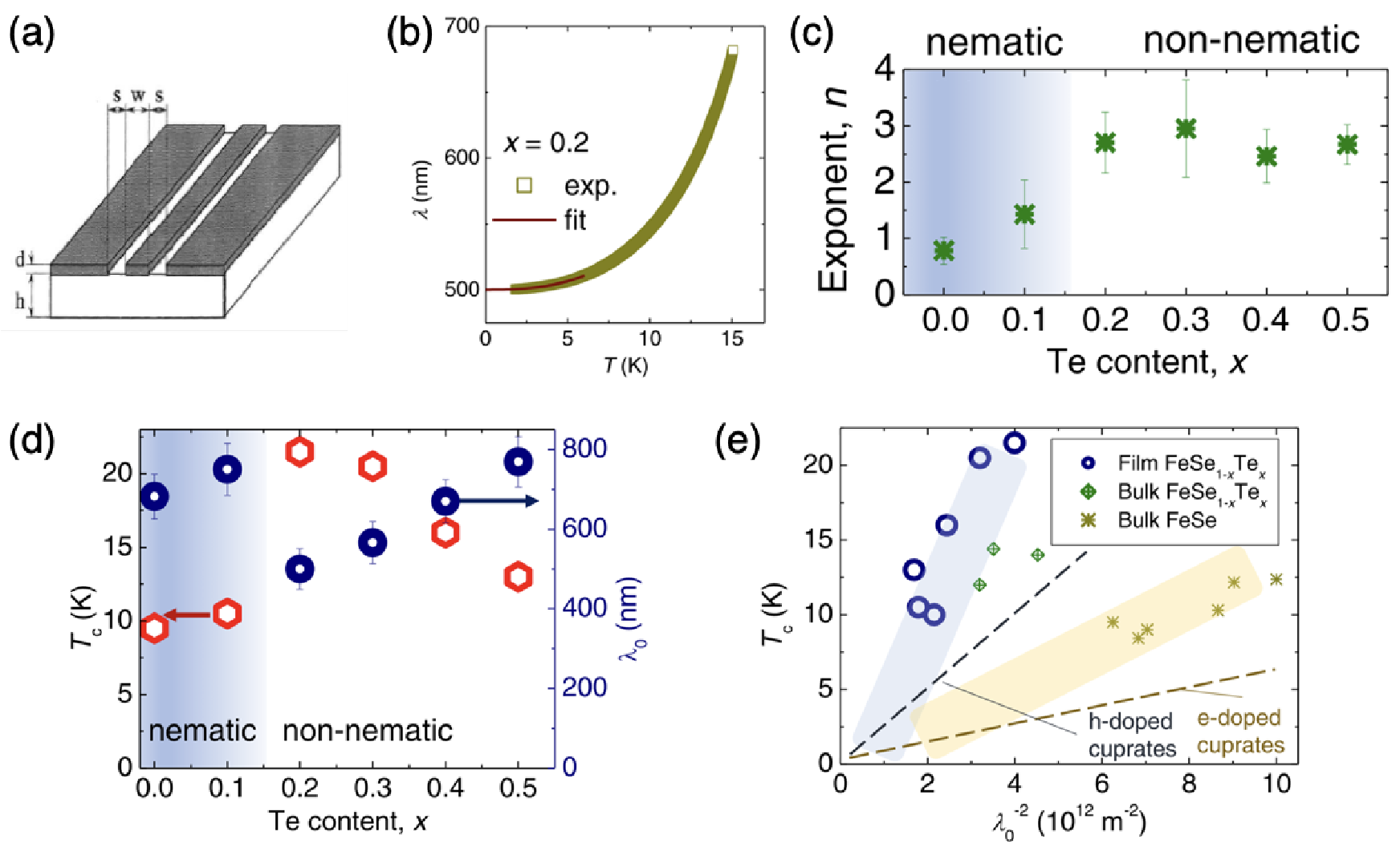}
\caption{Measurement of the magnetic penetration depth of a thin film sample of FeSe$_{1-x}$Te$_x$ on CaF$_2$ \cite{Kurokawa}. (a) Schematic diagram of a coplanar resonator.  In reality, the resonator part is meander-shaped. (b) Temperature dependence of the magnetic penetration depth of a sample with $x=0.2$.  The red line corresponds to a fitted curve with the equation $\lambda(T)=\lambda_0 + A(T/T_c)^n$.  (c) Composition dependence of the exponent, $n$, of the power-law temperature dependence. (d) Composition dependence of $T_c$ and the magnetic penetration depth at the low-temperature limit, $\lambda_0$. (e) $T_c$ as a function of $\lambda_0^{-2}$ in FeSe$_{1-x}$Te$_x$ ($x =$ 0$-$0.5) films. Results of bulk FeSe and bulk FeSe$_{1-x}$Te$_x$ ($x >$ 0.5) are also shown.  The blue line and yellow dashed lines correspond to the data of hole-doped cuprates and electron-doped cuprates, respectively.
Hatched areas are guides for the eye.  For references of bulk data and cuprate data, see descriptions in \cite{Kurokawa}.
}
\label{resonator}
\end{center}
\end{figure}

\

The charge dynamics of a superconductor is represented by the complex conductivity $\tilde{\sigma}=\sigma_1 + i\sigma_2$. Its behavior is precisely expressed by microscopic theory, but it is convenient to think of it as a two-fluid model that consists of a superfluid and a normal fluid, except near $T_c$.
In this model, the real part of the conductivity represents the dynamics of quasiparticles, and the imaginary part is mainly proportional to the superfluid density.
Therefore, investigating the temperature dependence of $\tilde{\sigma}$ can help infer the structure of the superconducting gap.
Specifically, for one type of carrier, in the low frequency limit, at frequency $\omega$ and temperature $T$,
\bea
\sigma_1(\omega,T)&=&\left(\frac{(1-f(T))ne^2\tau(T)}{m^*}\right)\frac{1}{1+\omega^2\tau(T)^2}, \label{QP}\\
\sigma_2(\omega,T)&=&\frac{f(T)ne^2}{m^*}\frac{1}{\omega},
\eea
where $n$, $m^*$ are the total carrier density and effective mass, $f(T)$ is the fraction of superfluid, which should be zero at $T_c$ and unity at $T = 0$ K if superconducting fluctuation is not taken into account, and $\tau(T)$ is the relaxation time of the quasiparticles.

Rather popular technique to measure $\tilde{\sigma}$ is the cavity perturbation technique in microwave and millimeter-wave region\cite{method}.
In this method, a small piece of crystal is put in the cavity and changes in the resonant characteristics, the resonant frequency,
$f$, and the quality factor of the cavity, $Q$, from which the complex conductivity $\tilde{\sigma}$ is obtained via the surface impedance $Z_s\equiv R_s-iX_s$ ($R_s$ and $X_s$ are the surface resistance and the surface reactance, respectively), since $Z_s$ is related to $\tilde{\sigma}$ as
\be
Z_s=\left(\frac{-i\omega\mu_0}{\tilde{\sigma}}\right)^{1/2}.
\label{Zs}
\ee

However, when this method is applied to a thin film sample, the electromagnetic field distribution around the sample changes significantly above and below $T_c$. In particular, in the low temperature limit, the ac magnetic field becomes essentially parallel to the thin film surface, which corresponds to the demagnetization factor being approximately unity for a thin film sample.\cite{Barannik} For this reason, it is not possible to calculate the surface impedance in all temperature ranges by a single common formula.
In addition, due to multiple reflections in thin film samples, the impedance measured is not the intrinsic surface impedance of the sample $Z_s$, but the effective surface impedance $Z_{eff}$\cite{Klein,Drabeck}.
\be
Z_{eff}=-\frac{i}{2}Z_s\cot\left(\frac{\omega\mu_0 d}{2Z_s}\right),
\label{nusol}
\ee
where $d$ is the thickness of the sample. Taking these circumstances into consideration, we adopted the combination of the following three methods.

(A) Penetration depth measurement of film samples processed into a strip to create a coplanar resonator (Fig.~\ref{resonator}~(a)),

(B) An ordinary cavity perturbation measurement in microwave magnetic field.

(C) A cavity perturbation measurement in microwave electric field.

Although the method (C) is usually used for nonconductive materials\cite{Peli1,Peli2}, I will show that the combination of (A)-(C) works very well for superconducting thin films.
Since this article is a comprehensive review, I will even skip important details of technical parts.
For those details, see \cite{Kurokawa,Matsumoto}.

\

\begin{figure}[htb]
\begin{center}
\includegraphics[scale=0.6]{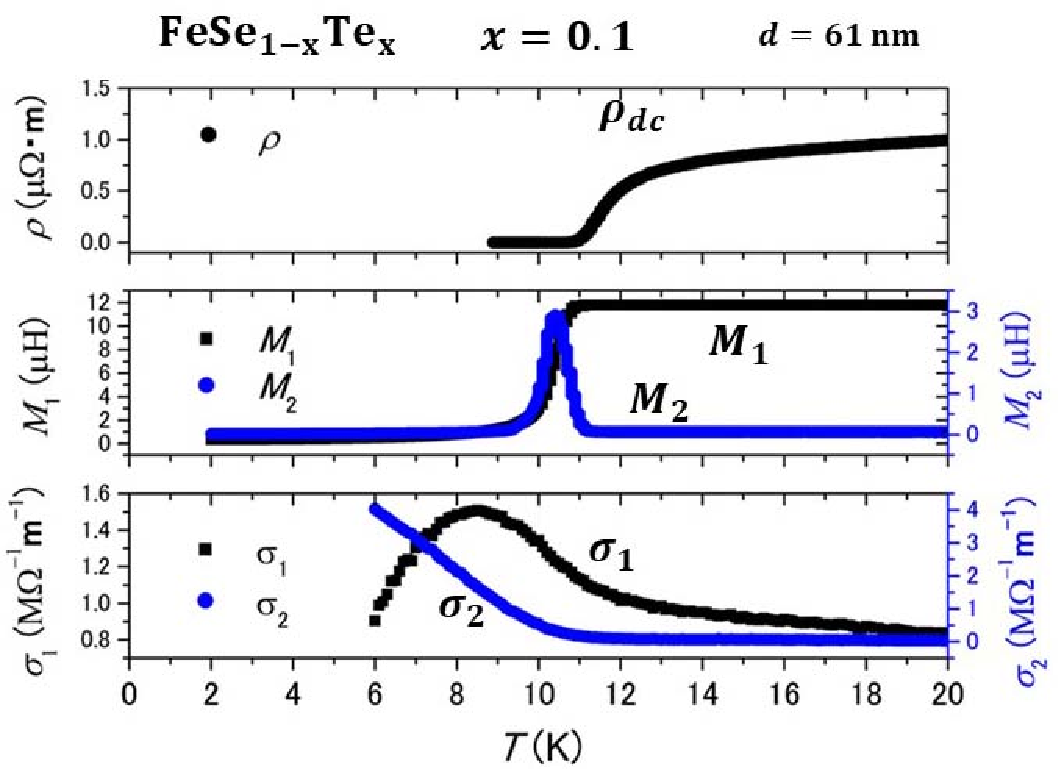}
\caption{Temperature dependence of  (a) dc resistivity, (b) ac magnetic susceptibility (the real part, $M_1$ and the imaginary part, $M_2$) at 33 kHz, and (c) microwave complex conductivity (the real part, $\sigma_1$ and the imaginary part, $\sigma_2$) at 44 GHz measured in microwave electric field of the same film sample.
All of these are for an FeSe$_{0.9}$Te$_{0.1}$ film with the thickness $d=$61 nm.\cite{Matsumoto}
}
\label{condE}
\end{center}
\end{figure}

Figure~\ref{condE} shows the comparison of the temperature dependence of dc resistivity, ac magnetic susceptibility at 33 kHz, and microwave complex conductivity at 44 GHz measured in microwave electric field of the same film sample.  Correspondence is rather good, and the microwave conductivity behavior is in good agreement with those obtained in bulk crystal qualitatively and even quantitatively\cite{Takahashi}, representing the measurement in microwave electric field works well.
It is also confirmed that the experimental results obtained by above mentioned two different methods connect well with each other\cite{Matsumoto}

\subsubsection{Superfluid density}

\

Figure~\ref{resonator}~(b) shows the temperature dependence of the magnetic penetration depth at low temperatures for the sample with Te substitution amount $x =$ 0.2.
It is expressed by a power law of temperature,
\be
\lambda(T)=\lambda_0+A(T/T_c)^n
\ee
 and the value of the exponent is a non-integer.
Such a non-integer power law is a phenomenon specific to multiband systems\cite{Mishra,Prozorov}. 
The measurement results of other samples are also expressed by the non-integer power law.
When the exponent n and the coefficient $A$ are plotted as a function of the Te content, as shown in Figure~\ref{resonator}~(c), the exponent changes before and after the composition at which the nematic transition disappears.
In the nematic state, n $=$ 1.0$\sim$1.5, whereas in the non-nematic states, n $=$ 2.5$\sim$3.0.
The former suggests a node-containing or highly anisotropic gap, while the latter suggests a nodeless or weakly anisotropic gap structure. 
Thus, the results suggest that the gap structure changes before and after the nematic transition.
The value of the penetration length $\lambda_0$ at the low temperature limit also changes before and after the transition,
As shown in Figure~\ref{resonator}~(d), it is longer when $T_c$ is low and shorter when $T_c$ is high.
However, they were found to obey the so-called empirical Uemura relation\cite{Uemura,Uemura2}
\begin{equation}
T_c \propto \lambda_0^{-2} 
\end{equation}
regardless of the nematic transition (Fig.~\ref{resonator}~(e)).

Recently, Prof. Yihua Wang's group at Fudan University performed a detailed study of the superconducting state of category 1 SC of FeSe$_{1-x}$Te$_x$ films\cite{KYLiang2025} which are combinatorially fabricated\cite{Qihong}.
They investigated local response of a narrow area of the film by scanning SQUID microscope, corresponding to the composition step of $\Delta x=$0.0008 and found that below and above the critical Te doping $x_c\simeq$ 0.23 which divides nematic states and non-nematic states, the SC is also divided into two different states.
For $x< x_c$ the SC gap is anisotropic and $T_c$ scales with phase stiffness whereas for $x>x_c$ the gap is isotropic and $T_c$ is determined by the gap amplitude with the maximum amplitude at $x_c$.
They also investigated the $T_c$ {\it vs} the superfluid density and found the so-called boomerang effect\cite{Uemura2}.
The presence of the nematic boundary at $x_c\simeq$0.23, anisotropic gap for $x<x_c$, isotropic gap for $x>x_c$, and the boomerang effect, all of these are in good agreement with already introduced results bove\cite{Imai1,Kurokawa,Matsumoto}.
However, based on their data, they concluded that whole features are indicative of the SC caused by nematic quantum fluctuation with $x_c$ as the nematic QCP.
The difference in the interpretation of whole features is closely related the different $T_c(x)$ relation between ref.~\cite{Imai1} and ref.~\cite{Qihong}.




\subsubsection{Dynamics of  quasiparticles}

\begin{figure}[htb]
\begin{center}
\includegraphics[bb=4 0 839 299,scale=0.5]{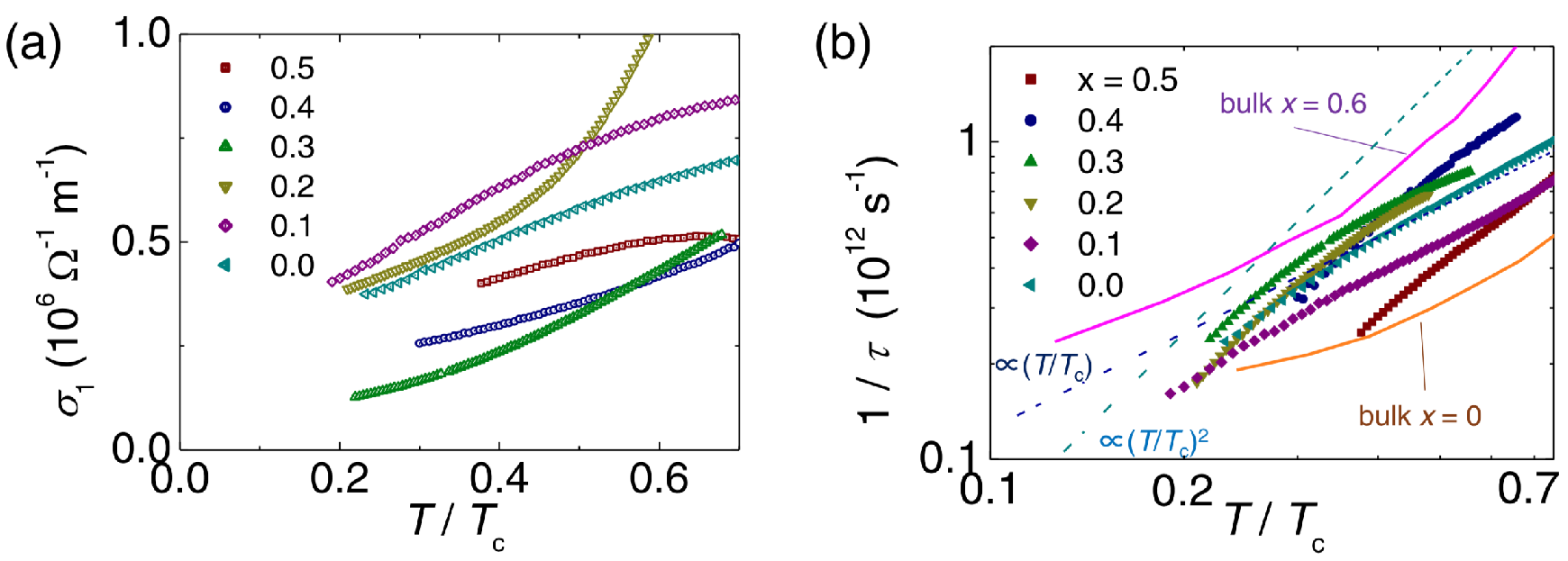}
\caption{ (a) Temperature dependence of the real part of the electrical conductivity for thin-film samples of FeSe$_{1-x}$Te$_x$ on CaF$_2$ with various Te content, $x$. (b) Temperature dependence of the relaxation rate of quasiparticles (the inverse of the relaxation time) for samples of various Te content. The results of bulk crystals are also shown as solid curves. \cite{Kurokawa}
}
\label{QPfig}
\end{center}
\end{figure}

\

Figure~\ref{QPfig} shows the temperature dependence of the real part of the electrical conductivity ((a)) and the reciprocal of the relaxation time of the quasiparticle (relaxation rate) $1/\tau$((b)) calculated by eq.~(\ref{QP}).\cite{Kurokawa}.
In FeCh, it is well known that multiple carriers are involved in the superconductivity of category 1.
However, here we will assume a representative carrier with the longest relaxation time and proceed with the analysis in a single band.

At all temperatures, $\tau$ increased, so $1/\tau$ was found to decrease below $T_c$.
The increase in $\tau$ when the material enters the superconducting state is a well-known behavior in cuprate superconductors, where the superconducting transition occurs under conditions where inelastic scattering is dominant\cite{Bonn}, and has also been observed in bulk single crystals of FeCh\cite{Takahashi,Okada}.

At low temperatures, the relaxation rate is expressed as  a power law in temperature,
\begin{equation}
   1/\tau = aT^k + b.
\end{equation}
Comparing the results for different Te contents,
we found that in the nematic state, the exponent is $k\sim 1$, whereas
in the non-nematic state, the exponent is $k> 2$.
In comparison with previous discussions on several superconductors, including iron-based superconductors\cite{MLi,Hirschfeld,Ozcan,Hashimoto,Quinlan}, which discussed the relationship between the exponent $k$ and the gap structure,
we found that the exponent of $k \simeq 1$ in the nematic state corresponds to an anisotropic gap,
and the exponent of $k > 2$ in the non-nematic state corresponds to nodeless gap.
This picture obtained for the quasiparticle dynamics is consistent with the results of the superfluid density measurement mentioned above.

In contrast to the sharp change in the SC gap structure at the nematic boundary observed in film samples, the behavior in bulk crystals is rather different and complicated\cite{bulkQCP}.
However, the data in Fig.~\ref{phaseex} suggest all data including bulk, compressively strained film on LAO, and compressively strained film on CaF$_2$, look systematic. in terms of $T_c$ as a function of Te content.
Remembering that in thin films,
a pure nematic transition, which is a nematic transition
that does not involve lattice deformation, occurs\cite{Kubota},
it can be considered that the complicated behavior in bulk crystals might be interpreted as the coupling with lattice obscured the intrinsically sharp effect.
The complicated effect of the lattice was discussed in \cite{Labat2017}.

\subsubsection{Superconductivity fluctuation}

\

Before closing the superconductivity property section of category 1,
we discuss the superconducting fluctuations of FeSe.

It was pointed out that the Fermi surface volume is extremely small, comparable to superconducting gap, in FeSe by ARPES\cite{Lubashevsky2012, Okazaki2014} and STM\cite{Kasahara2014}.
Specifically, the ratio of the superconducting gap $\Delta_{SC}$ to the Fermi energy $E_F$ is $0.1\sim 1$.
This immediately suggests the presence of huge superconductivity fluctuation, since the barometer of superconductivity fluctuation, the Ginzburg number $G_i$, is represented as $G_i=80(T_c/T_F)^4$\cite{LarkinBK} which becomes $\sim 0.1$, in sharp contrast to that for conventional superconductors ($\sim 10^{-4}$)\footnote{This is an expression for clean 3 dimensional superconductor.  Since Fe chalcogenides has anisotropic, expression should be modified.  However, the dominant factor is some power of $T_c/T_F$, representing he essential feature does not differ so much.}.

Magnetic torque study using Piezo element suggests that the ratio of the onset of superconductivity fluctuation $T^*$ to $T_c$ is huge, $T^*/T_c\sim 2.35$\cite{Kasahara2016}.
On the other hand, another magnetic torque study using optical interference by ourselves suggests $T^*/T_c\simeq 1.1\sim 1.2$\cite{Takahashi2019}.
Magnetic susceptibility study suggests $T^*/T_c\sim 1.03$\cite{Yuan2017}, whereas an NMR study suggests $T^*/T_c\sim 1.76$\cite{Shi2018}.
All of these are for bulk crystals.

In PLD grown films, superconductivity fluctuation seen in complex conductivity measured by microwave broadband technique\cite{Booth,KitanoRSI}, specific technique to film samples, showed that the fluctuation is seen at most up to 1.2 $T_c$ ($T^*/T_c\sim 1.2$)\cite{Nabeshimafluc}, which agrees well with the magnetic torque result by ourselves\cite{Takahashi2019}.
The complex conductivity study also at microwave frequency (44 GHz) in microwave electric field\cite{Matsumoto} shows that the temperature dependence of superfluid density is seen up to 1.2$T_c$, again in good agreement with \cite{Takahashi2019} and \cite{Nabeshimafluc}

Thus, although the superconductivity fluctuation is not so anomalously large, it is still large when compared with conventional superconductors.

It should be added that the enhancement of superconductivity by the optical pumping is observed in FeSe PLD films\cite{Isoyama2021}.
This counterintuitive observation also shows another characteristic aspect of superconductivity fluctuation in FeSe film.

\subsubsection{Summary of properties in the superconducting state}

\begin{figure}[htb]
\begin{center}
\includegraphics[scale=0.3]{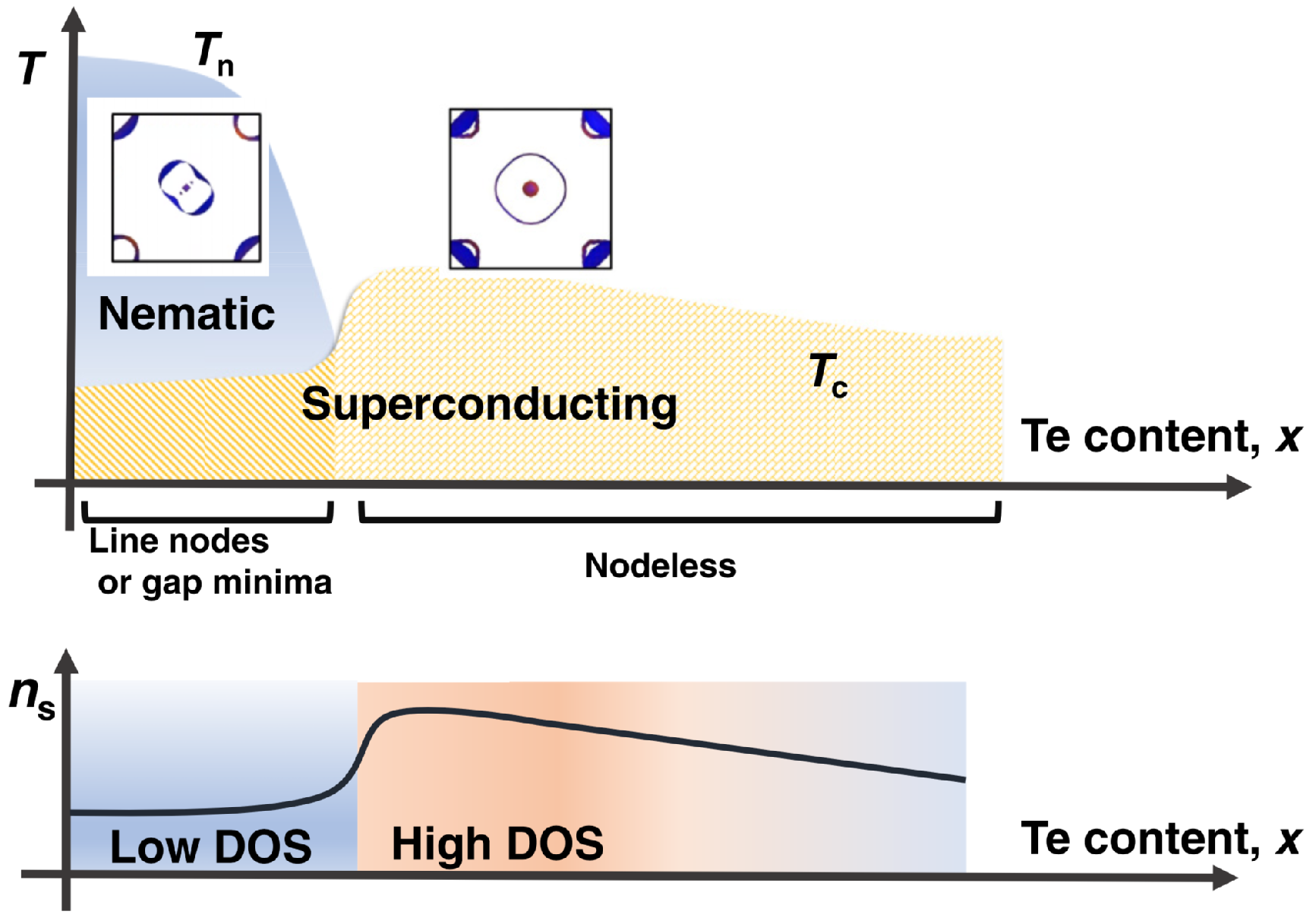}
\caption{Schematic diagram of the category-1 superconducting state of FeSe$_{1-x}$Te$_x$. (top) Composition dependence of the nematic transition temperature, $T_n$ and the superconducting transition temperature, $T_c$, represented by blue and yellow hatchings, respectively. (bottom) Te composition dependences of the superfluid density $n_s$ and density of statesare also shown as a solid curve and gradations, respectively.\cite{Kurokawa}.
}
\label{phaseSC}
\end{center}
\end{figure}

\

To summarize what was found on the category 1 superconducting state for a series of thin film samples with Te substitution,

(1) $T_c$, $B_{c2}$, {\it etc.} change significantly around the Te concentration $x_c$ at which the nematic order disappears\cite{Sawada}.
Measurements of the superfluid density and quasiparticle dynamics suggest that the gap structure in the superconducting state changes there.
Explicitly, it was suggested that there is a very anisotropic gap for $ x < x_c$, and a weakly anisotropic gap structure without nodes at $x > x_c$ (Figure~\ref{phaseSC}).

(2) Nevertheless, it was found that $T_c$ is proportional to the superfluid density $n_s$ regardless of the presence or absence of nematic order, suggesting that the most important factor determining $T_c$ is not the presence or absence of nematic order, but the carrier density.
Indeed, $T_c$ had a positive correlation with the carrier density in the normal state.
From these, it can be said, although the structure of the gap will be affected by the details of the electronic state, the gross mechanism of pairing itself might be robust and is unaffected by the details of band structure.

(3) Different phase diagrams $T_c(x)$ have been proposed even among PLD grown films, that remain above mentioned statements open to further studies.

\section{Category 2}

Now, let us move on to category 2. 
Since the discovery of copper oxide high-$T_c$ superconductors, it has been widely recognized among physicists that carrier doping on the order of a few\% to 10\% is the key to controlling physical properties.
However, a breakthrough that succeeded in increasing the doping amount by an order of magnitude was carrier doping by the electric field effect using the electric double layer\cite{Ueno}.
This method was quickly tried for iron-based superconductors, especially FeCh, and it was reported that the $T_c$ could be increased by carrier doping, and that SC of 40 K class could be achieved\cite{Shiogai,Hanzawa,Lei}.
From measurements of the Hall effect, it was confirmed that the carrier of this SC is on the electronic Fermi surface at the M point\cite{Hanzawa,Lei,Shiogai2}, and it can certainly be said to be category 2 superconductivity.
Furthermore, it was shown that electrochemical etching of the sample was possible while keeping the same electric double-layer transistor structure, which opened a new avenue for the fabrication of ultrathin films\cite{Shiogai,Shiogai2}.
In the report by Shiogai {\it et al.}\cite{Shiogai}, it was necessary for the sample thickness to be below a certain level to achieve category 2 SC.
On the other hand, our experiments using various substrates showed that there is no such ``critical film thickness."
This suggests that rather than electrostatic doping, some kind of reaction occurs in a region of about 10 nm deep on the surface, resulting in category 2 SC, at least in our case\cite{Kouno}.
By further investigating the doping process, we were able to reproducibly fabricate samples with zero resistance at 46 K on multiple different substrates, as shown in Fig.~\ref{EDLT}.
To our knowledge, this zero-resistance value is the highest in the world for iron chalcogenides\cite{Shikama1,Shikama2}, except for the singular report by Ge {\it et al.} \cite{Ge}.

Figure~\ref{phaseex} shows the effect of homologous substitution (chemical pressure effect) on the electronic surface-dominated category 2 SC realized by field-effect doping\cite{Shikama2}. 
For both S and Te substitutions, $T_c$ decreases.
The phase diagram is simpler because there is no nematic transition, and the chemical pressure effect on $T_c$ ia rather different form that in category 1.  Thus, it is natural to consider that the mechanism of SC is different between category 1 and category 2.
As for categorfy 2 behavior, it can be interpreted as follows.
The decrease of $T_c$ for both nagative and positive pressure is caused by the introduction of randomness\cite{Maekawa}.
In this sense, category 2 SC is closer to the ``standarad'' SC.

Jiang {\it et al.} realized a continuous variation of $T_c$ from 10 K to 40 K by EDLT doping on FeSe films prepared by the PLD technique, which probably belongs to ``category 2'' superconductivity\cite{Jiang2023}.
However, they did not observe any discontinuous change of $T_c$ as a function of doping.
In all samples, resistivity changes linear in temperature close to $T_c$ in the normal state.
They found a positive correlation between $T_c$ and the temperature coefficient of $T$-linear resistivity, which is very similar to that found in cuprate superconductors.
Based on these, they concluded that the pairing mechanism is suggested to be anti-ferromagnetic fluctuations.
Indeed, they explained the behavior in term s of a cooperative effect of quantum fluctuation and disorders\cite{Zhang2025}.

\section{Category 3}

\begin{figure}[htb]
\begin{center}
\includegraphics[scale=0.5]{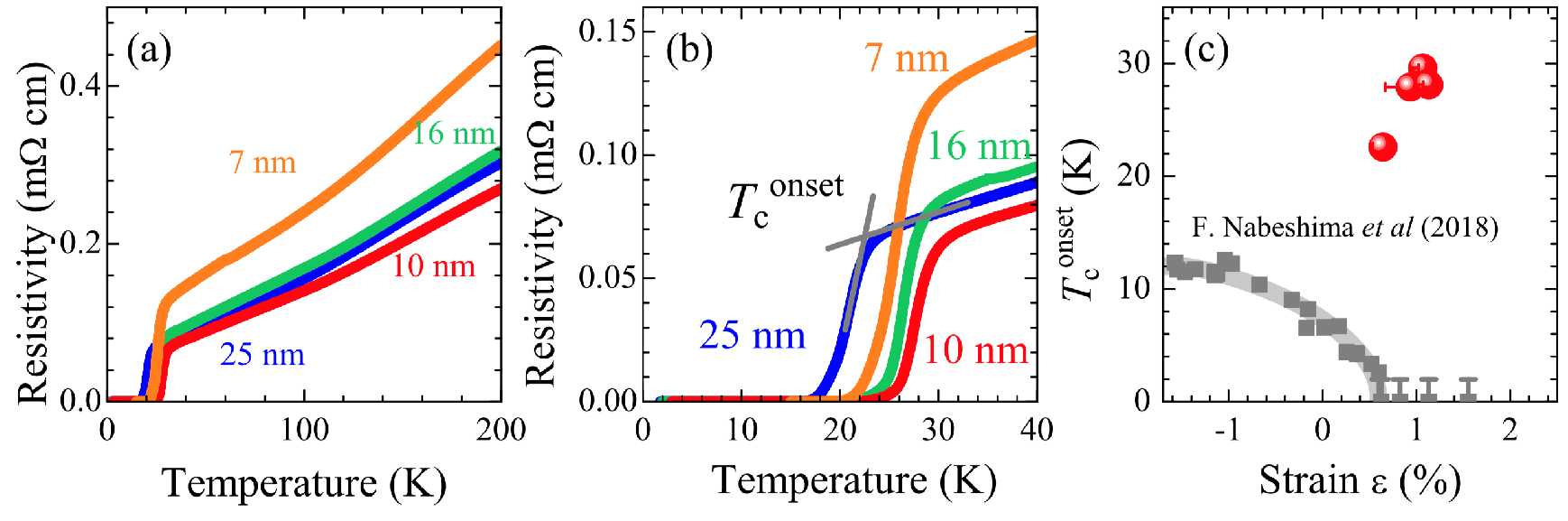}
\caption{(a) Temperature dependence of electrical resistivity of FeSe thin film samples on STO, (b) enlarged view in the vicinity of $T_c$, (c) Red balls show $T_c^{onset}$ of these samples plotted on a strain vs. $T_c$ diagram.  Gray circles are data taken from ref. \cite{Nabeshima2}.  A gray curved band is a guide for the eye.(\cite{Kobayashi})   
}
\label{interface}
\end{center}
\end{figure}

All research on SC in category 3, which is the most interesting for FeCh, has been conducted using samples deposited by MBE. If monolayer films can be produced by the PLD method, it will have some advantages over MBE.
(a) We do not need as much active oxygen as MBE when depositing oxide films.
(b) By changing targets one after another, material search and interface design will be easier than with MBE, and research on interface effects on superconductivity will be expanded.

Since the PLD method inevitably generates droplets, it may seem impossible to form films at the atomic layer level. However, layer-by-layer film formation at atomic layer level has been performed for oxides\cite{LASER-MBE}, which is also called laser MBE\cite{Pascal}. Therefore, it cannot be said that the PLD method is in principle unsuitable for forming films at the atomic layer level. It is only a matter of compatibility with the target material.

As mentioned in the previous section, Shiogai {\it et al.}\cite{Shiogai} claim to have succeeded in forming a film at the atomic layer level by etching, starting from a thin film sample formed by PLD. However, since they estimated the film thickness by monitoring the amount of gate leakage current during etching, they cannot obtain information on the actual state of the film formation.
For example, it is difficult to imagine that a uniform monolayer of FeSe grows on MgO, which has a completely different lattice constant from FeSe. In fact, Shiogai {\it et al.} performed deposition and EDLT processes on various different substrates and compared the superconducting properties, and reported that there is a universal relationship between $T_c$ and the Hall coefficient regardless of the type of substrate\cite{Shiogai2}, suggesting that this superconductivity is not category 3 as defined in this paper, but category 2.
However, detailed temperature and angle dependence of the critical magnetic field, including anisotropy, were measured for samples with various film thicknesses\cite{Shiogai3}, and it was found that the GL coherence length perpendicular to the film was less than 1 nm, and that the coupling between the EDLT-doped layer and the doped layer from the substrate plays an important role in interpreting the results.

In contrast, Hirahara group\cite{Pedersen}, which has realized superconductivity in a few atomic layers by MBE,
focused from the beginning on the effect of differences in the surface structure of the STO substrate on the superconducting properties of the film. By comparing the superconducting properties of various surface structures, they obtained the result that the size of the gap as seen by STM as a function of the carrier concentration doped from the substrate draws a dome structure, which is strong evidence of the interface effect on SC\cite{Pedersen,Tanaka0,Tanaka1,Tanaka2}.

\begin{figure}[htb]
\begin{center}
\includegraphics[scale=0.25]{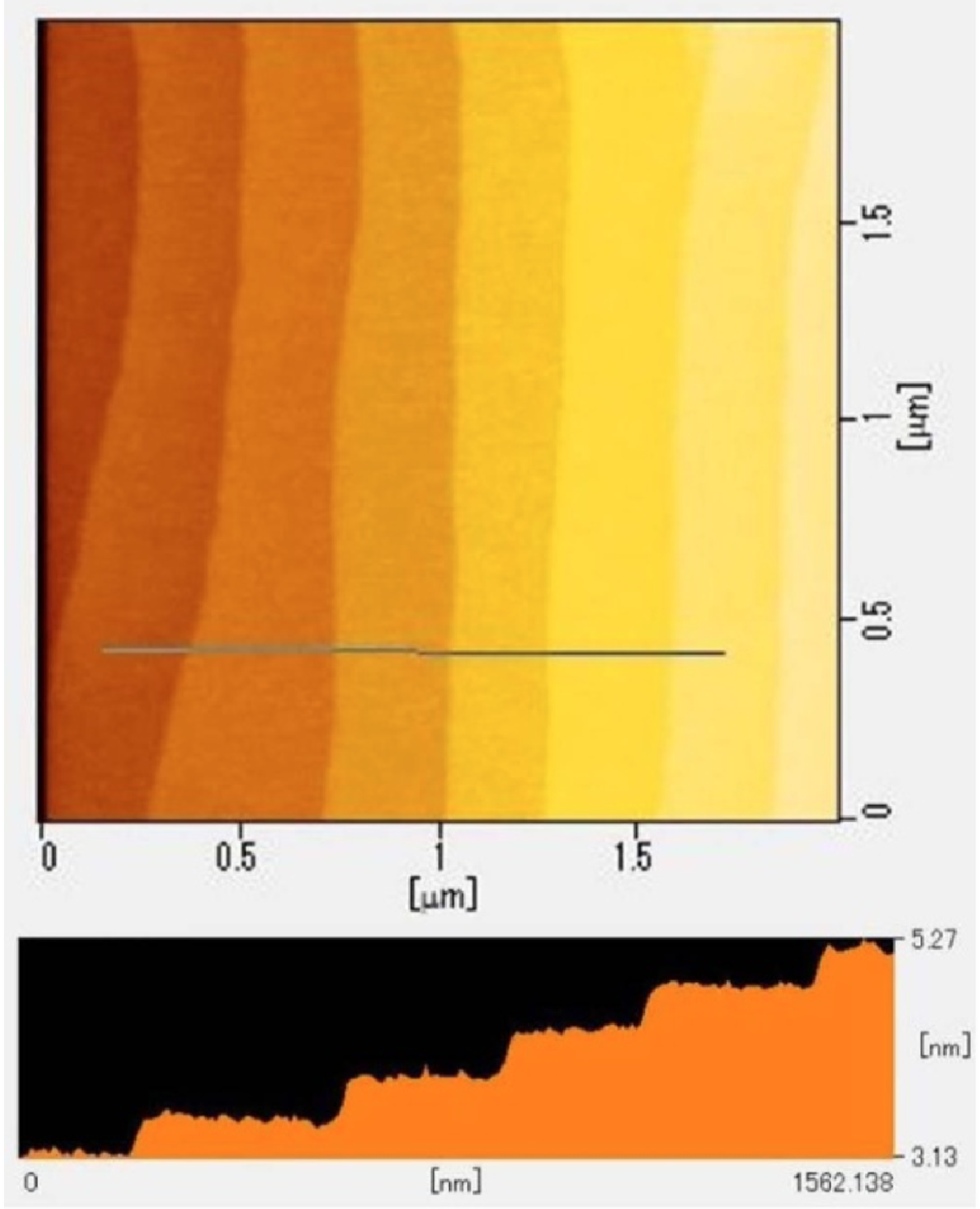}
\caption{Atomic force microscope topographic image of an STO substrate after the surface treatment. Clear step terrace structures are
observed\cite{Kobayashi}.
}
\label{step}
\end{center}
\end{figure}

\begin{figure}[htb]
\begin{center}
\includegraphics[,scale=0.6]{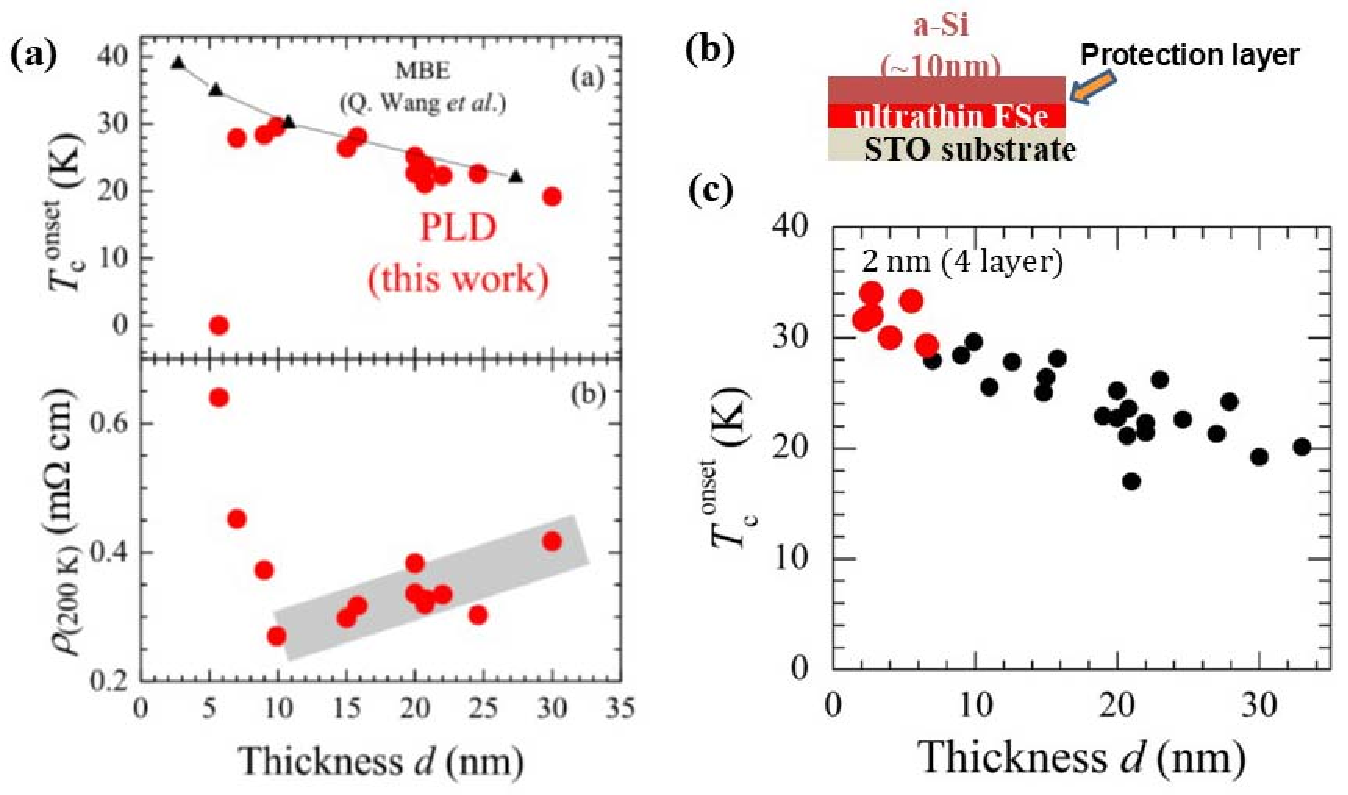}
\caption{(a) (top) Thickness dependence of $T_c^{onset}$ for FeSe thin films on STO (red circles). Results by MBE are also shown by black triangles.  (bottom) Thickness dependence of electrical resistivity at 200 K. A gray band is a guide for the eye\cite{Kobayashi}, (b) Schemtic figure of protection capping of FeSe film. (c) improved data of $T_c$ after the protection capping as a function of film thickness.
Black circles and red circles represent the data without protection capping and those with protection capping, respectively. 
}
\label{thickness}
\end{center}
\end{figure}

Thus, keeping in mind that the structure of the substrate surface is important, we started with films with normal thickness and gradually approached ultrathin films by PLD, examining the deposition conditions. We fabricated samples on STO with thicknesses ranging from 30 nm to 2 nm, and all of them showed a superconducting transition with an onset $T_c$ of about 30 K (Fig.~\ref{interface}~(a)(b)) \cite{Kobayashi}. 
The magnitude of strain can be evaluated by measuring the in-plane lattice constant, and when we plotted the relationship between strain and $T_c$ discussed in {\it sec. 2.2}, we got Fig.~\ref{interface}~(c), which shows that the relationship between strain and $T_c$ is completely different from the established relationship\cite{Nabeshima2}.
We believe that the reason for the large deviation is the effect of the surface treatment of the substrate. In other words, the sample that became superconducting here was deposited after confirming that the so-called step-terrace structure (Figure~\ref{step}) appeared on the substrate surface due to the surface treatment of the substrate. 
In this way, since the sample with the surface treatment of the substrate showed results that deviated from the conventional relationship between strain and $T_c$, it is believed that the effect of the interface on SC is also observed here.
This is supported by the results of the film thickness dependence of $T_c$ shown in Fig.~\ref{thickness}. 
As can be seen from this figure, $T_c$ increases as the film thickness decreases.
This is qualitatively and even quantitatively identical to the ultrathin film samples fabricated by MBE that claim to have an interface effect\cite{Wang2}. 
Thus, for the first time, we have succeeded in realizing the interface effect on the superconductivity of FeCh by the PLD method.
In the early state\cite{Kobayashi}, the increase in $T_c$ stopped when the film thickness was 5 nm, which was found to be because of  the surface deterioration due to the air exposure.
We now put protection layers (Fig.~\ref{thickness}~(b)), and have confirmed superconductivity down to $t =$ 2 nm (about 4 layers) (Fig.~\ref{thickness}~(c)).

\begin{figure}[htb]
\begin{center}
\includegraphics[scale=0.5]{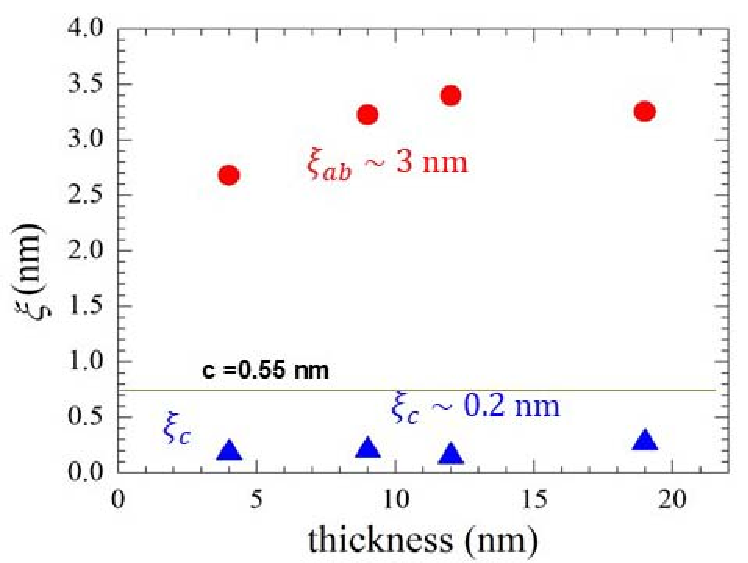}
\caption{Coherence length estimated by the uppercritical field measurment for various films with different thickness.
A horizontal solid line represents the $c-$axis length of FeSe (after\cite{Kobayashi2}).}
\label{figcoherence}
\end{center}
\end{figure}

Another supporting evidence of interface superconductivity is the coherence length estimated by the upper critical field measurement\cite{Kobayashi2}.
As shown in Fig.~\ref{figcoherence}
In-plane $\xi_{ab}$ is about 3 nm, in agreement with the MBE film data\cite{Wang3}.
On the other hand, the coherence length in the perpendicular direction, $\xi_{c}$ is $\sim$ 0.2 nm, which is less than one layer  length of FeSe, and is independent of the total film thickness.
This is a representation that SC occurs in the confined area at the surface between the substrate and FeSe.

Very recently Zhao {\it et al.} observed essentially the same superconductivity in about 20nm-thick FeSe films grown by the PLD technique on ST0, with $T_c^{onset}$, $T_c^{zero}$ of 28.3 K and 23 K, respectively.
The $I-V$ characteristic investigation shows the BKT features with $T_{BKT}$ of 17.8 K.
Magneto-transport experiment showed that carrier density does not differ so much from that in ``category 1'' FeSe, and the obtained effective mass is large, with which they concluded that the correlation effect is important for the realization of ``high-$T_c$'' superconductivity in this system.
If superconductivity occurs at (or around) the interface, transport properties is the summation of that at the interface and that at the different part of the films.
Indeed, the Nernst data clearly shows that both the ``category 1'' part and the interface part contribute the observed Nernst signal\cite{Kobayashi3}.
Thus, it is not clear whether the observed transport properties (and also the ARPES data in the same literature) by Zhao {\it et al.} come from the interface part contributing the superconductivity of current interest or not.
\begin{figure}[htb]
\begin{center}
\includegraphics[scale=0.6]{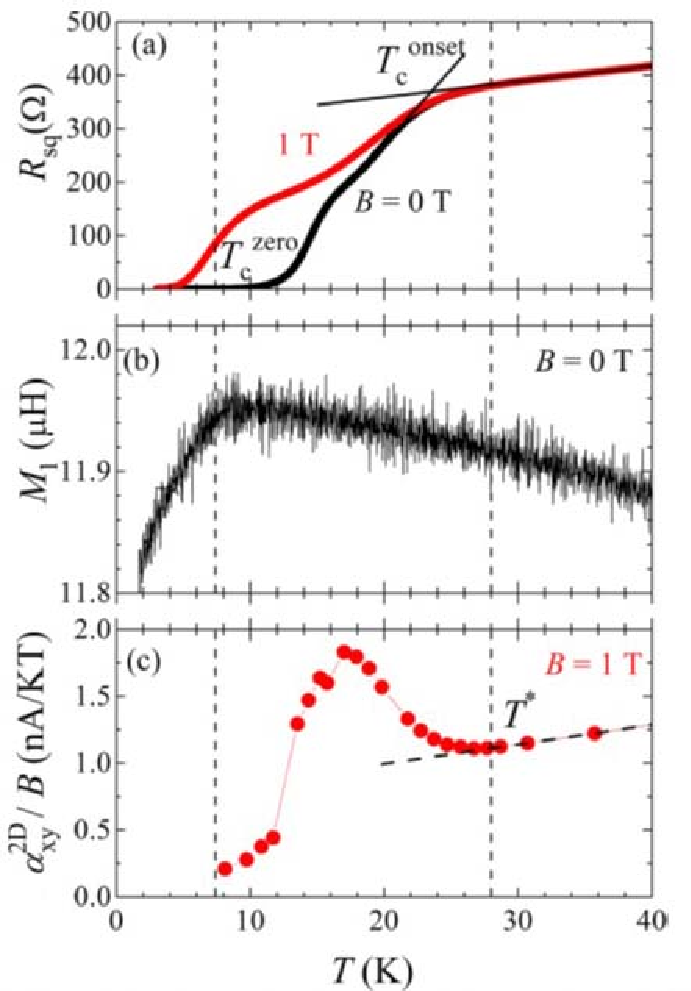}
\caption{The comparison of (a) dc sheet resitance, $R_{sq}$, (b) magnetic susceptibility (real part of mutual inductance $M_1$), and (c) Nernst effect $\alpha_{xy}^{2D}/B$ of the same film sample as functions of temperature.  Dashed lines represenrt $T_c$ determined by diamagnetism and by the Nernst effect respectively.\cite{Kobayashi3}
}
\label{compare}
\end{center}
\end{figure}

Anyway, the interface SC in FeSe/STO is realized even by the PLD technique.
Then, our most fundamental question; from what temperature does SC start?
Figure~\ref{compare} shows the comparison of dc resistivity, magnetic susceptibility measured by mutual inductance technique, and Nernst effect of the same film sample as functions of temperature\cite{Kobayashi3}.
In this sample, large diamagnetic signal appeared almost at the zero-resistance temperature, whereas the Nernst signal related to SC shows up from slightly higher temperature that the onset $T_c$ of SC shown in resistivity.
However, it is still at most about 30 K.

Among the diamagnetism data published so far, the data in refs.\cite{diamag1,WZhang2014,YSun2014} are similar to the above presented data in the sense that large diamagnetic signal appeared almost below the zero-resistance temperature, which is in sharp contrast to the previously mentioned singular reports\cite{diamag2,YSong2021}.
\footnote{The authors in ref.\cite{YSun2014} 
describe the possibility of 85 K as an onset of diamagnetism related to SC.  However, unbiased look at the data suggests that the onset of diamagnetism related to SC is around zero-resistance temperature.}

Thus, it has not yet been possible to stably realize category 3 superconductivity in terms of electrical resistance.

\begin{figure}[htb]
\begin{center}
\includegraphics[scale=0.6]{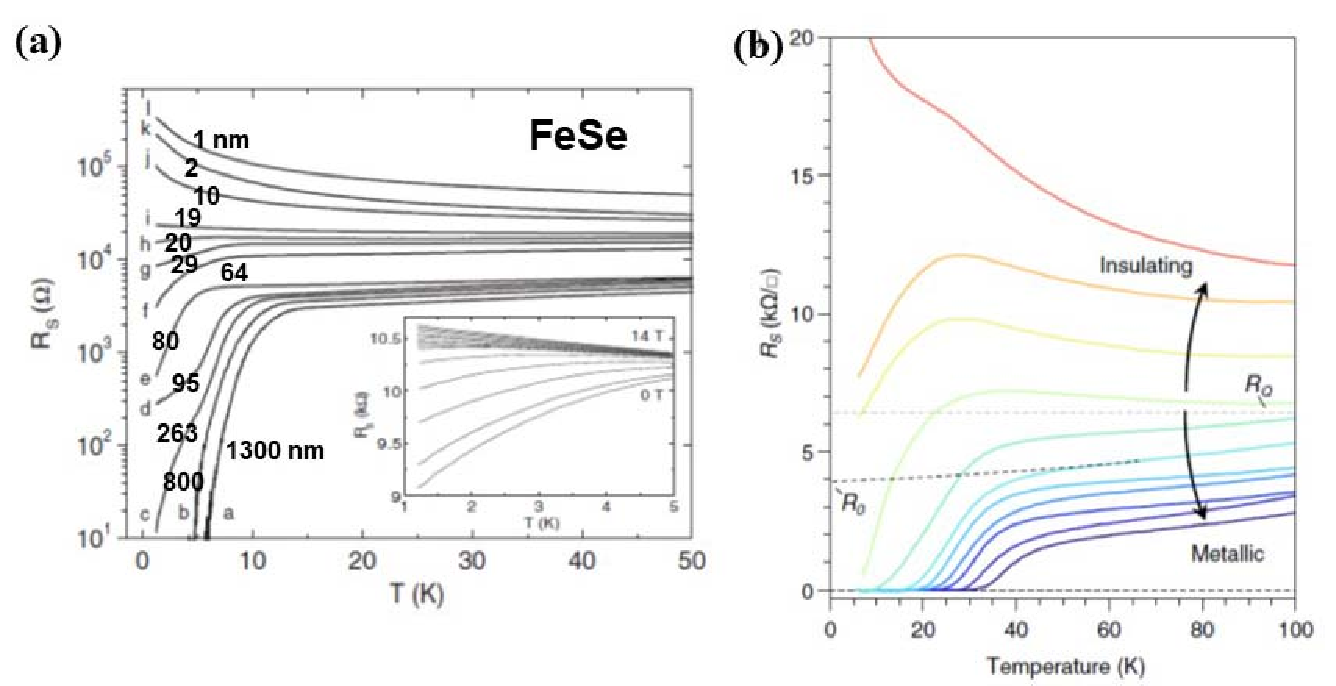}
\caption{Superconductor-insulator transition in FeSe. (a) early data of films fabricated by sputtering with varying sample thickness\cite{Schneider}. Numbers representing film thickness are added by us.  The inset shows the sheet resistance $R_s$ a 30 nm thick film as a function of temperature for various magnetic field as a parameter. The values of B are 0, 1, 3, 5, 7, 8.5, 8.9, 9.3,
9.8, 10.5, 11, 12, 13, and 14 T (from bottom to top). (b) more recent data of MBE monolayer films annealed under different conditions. For details on the conditions, see the original reference\cite{Faeth}.
}
\label{figSIT}
\end{center}
\end{figure}

For SC studies in films, one of the important viewpoints is the issue of the superconductor-insulator transition (SIT).
When a superconductor sample containing a certain amount of randomness is thinned, the SIT occurs when the sheet resistance reaches the pair quantization resistance $R^{crit}_{sq} = h/(2e)^2 =$ 6.45 k$\Omega$ ($h$ and $e$ are the Planck constant and the elementary charge, respectively)\cite{MPA,SIT,SITr}.
In fact, there have already been reports of the SIT for FeSe.
Figure~\ref{figSIT}~(a) shows the temperature dependence of sheet resistance of FeSe films fabricated by sputtering
where the thickness of the sample is changed from 1300 nm to 1 nm, and the results show that the SIT occurs at a thickness of about $t =$ 20 nm, with the resistance at this boundary being approximately 6.45 k$\Omega$\cite{Schneider}. 
On the other hand, in an experiment conducted by the Cornell University group in which the annealing conditions were changed for monolayer MBE FeSe films to systematically change the disorder, SIT was observed roughly at $R^{cirt}_{sq}$\cite{Faeth}, representing the improvement of the film quality.

In the case of our data shown in Figs~\ref{interface} and \ref{thickness}, for the $t =$ 20 nm film and $t =$ 10 nm film, $R_{sq} \simeq$ 0.35 m$\Omega$cm/20 nm = 175 $\Omega$ and $R_{sq} =$ 0.27 m$\Omega$cm/ 10 nm = 270 $\Omega$, respectively, both of which are significantly smaller than $R^{crit}_{sq}$.
In fact, for both films, the temperature dependence of the resistance in the normal state is metallic, and a superconducting transition is also observed.
Thus, the film thickness correspoding to $R^{crit}_{sq}$ should be much smaller than that in Fig.~\ref{figSIT}~(a). showing high quality of our films.

Recent research on two-dimensional superconductors has suggested that in clean two-dimensional superconductors with little disorder, there is a metallic phase between the superconducting phase and the insulating phase for various materials\cite{Saito,SITr,Ienaga}. This phase is characterized by a metallic sheet resistance much smaller than the above $R^{crit}_{sq}$ and a zero Hall resistance and is therefore called a "Bose metal" state. In fact, as mentioned at the beginning, such a state has been reported for ultrathin FeSe films\cite{YLi}.
Although no consensus has yet been reached, this is a subject of current active research.

A theoretical study has shown that during such SIT, islands of superconductor or insulator form in real space and change size with thickness.\cite{Swanson}.
The same paper also discusses the ac conductivity in each case.
Therefore, if we could prepare samples of various thicknesses and measure the local conductivity in addition to the usual ac conductivity, we would be able to discuss the possible domain formation and the validity of this theory. This would then be a starting point for a more detailed discussion of, for example, the extent to which the previous scheme of the SIT is valid for the category 3 superconductivity observed on oxide substrates.

\begin{figure}[htb]
\begin{center}
\includegraphics[scale=0.5]{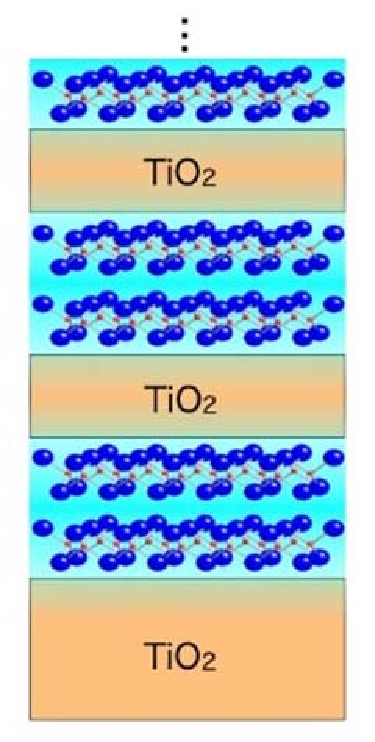}
\caption{Schematic illustration of the superlattice of FeCh with multiple interfaces between TiO$_2$ surface and FeCh.
}
\label{ideacite}
\end{center}
\end{figure}

If the reason for the low zero-resistance temperature is the BKT transition due to the ultrathin film\cite{BKT,BKT2}, can we expect the increase of the zero-resistance temperature by weakening the two-dimensionality?
In other words, can we kill two dimensionality by keeping the presence of the interface and by introducing the coupling between the interfaces? 
Based on this idea, we are trying to design and form superlattices and multilayers by oxide and FeCh, aiming to realize $T_c$ above 65 K with zero resistance (Fig.~\ref{ideacite}).
In fact, we have already experienced the fabrication of FeCh-based superlattices, and have succeeded in obtaining a high $T_c$ that cannot be obtained by a single composition\cite{NabeshimaSL}.
However, the present task is to fabricate superlattices between chalcogenides and oxides, which is more challenging.

\begin{figure}[htb]
\begin{center}
\includegraphics[scale=0.45]{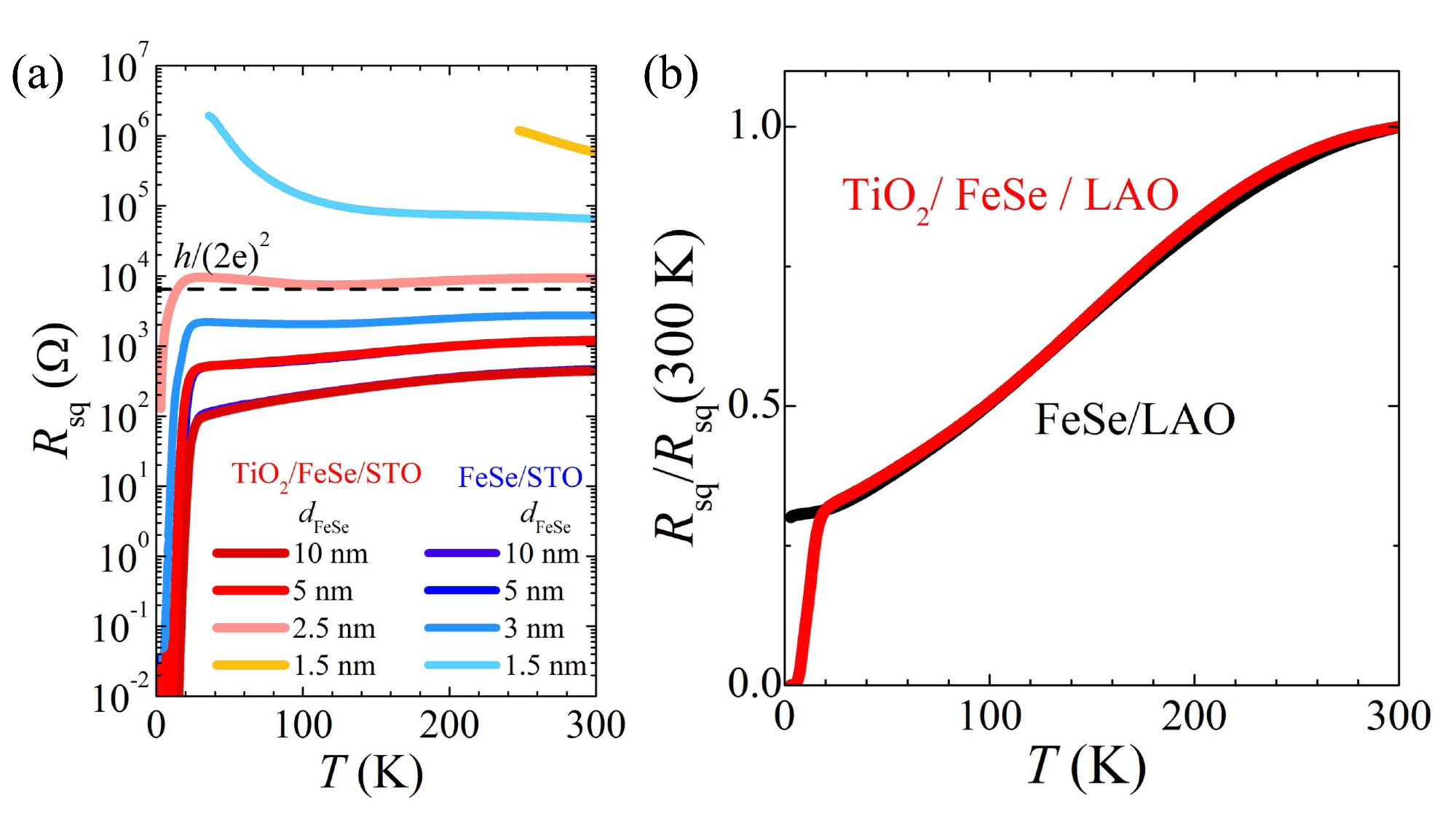}
\caption{(a) Temperature dependence of the sheet resistance of TiO$_2$/FeSe/STO (red, magenta and orange curves) and FeSe/STO (blue curves) films with different $d_{{\rm FeSe}}$. For TiO$_2$/FeSe/STO heterostructures, the TiO$_2$ thickness was kept at
d$_{{\rm TiO}_2}$$\sim$2 nm. (b) Temperature dependence of the normalized resistance of TiO$_2$/FeSe/LAO (red curve) and FeSe/LAO (black curve) films with d$_{{\rm FeSe}}$ $\sim$ 5 nm (taken from \cite{KobayashiNY}).
}
\label{SLtrial}
\end{center}
\end{figure}

Figure~\ref{SLtrial}~(a) shows resistivity of TiO$_2$/FeSe/STO with various FeSe thickness as a function of temperature, together with those without TiO$_2$ on FeSe\cite{KobayashiNY}.
Both structures have a-Si as a protection layer on the top.
For films thicker than 5 nm, both structures become superconducting, whereas for films thinner than 5 nm both do become insulating.
Thus, no difference was found between the structure with TiO$_2$ on FeSe and without TiO$_2$, and it is likely that deposition of TiO$_2$ on FeSe provides damages for thinner FeSe films.
Figure~\ref{SLtrial}~(b) shows resistivity of TiO$_2$/FeSe/LAO as a function of temperature, together with that without TiO$_2$ on FeSe.
In this case, the structure without TiO$_2$ does not become superconducting down to lowest temperature measured.
On the other hand, the structure with TiO$_2$ on FeSe shows superconductivity.
This represents that SC realizes at the interface of FeSe/TiO$_2$.
For both of Fig.~\ref{SLtrial}~(a) and (b), we have not succeeded in reaching thin-film hybrid structure where the coupling between the interfaces exists.
However, the successful realizaion of SC at the interface of FeSe/TiO$_2$ is promising.
This direction should deserve further investigation.

\section{Summary and Future Outlook}

We have reviewed the three categories of superconductivity that appear in iron chalcogenides FeSe$_{1-y}$S$_y$ and FeSe$_{1-x}$Te$_x$, focusing on studies using thin film samples, particularly those fabricated by PLD.
According to these studies, it seems plausible that the mechanism of superconductivity is different for the three categories, but it is not easy to identify the mechanism of superconductivity microscopically.
However, if we interpret the fact that a pure nematic transition was observed in the thin-film samples of category 1 and that the $T_c$ was higher than that of the bulk samples at ambient pressure because of the negligibly small lattice effect, it may not be unreasonable to think that phonons do not play a dominant role in pair formation.

What is of most interest is the stable realization of Category 3 high-temperature SC by electrical resistance, which has not appeared in front of us yet.
Even we do not really know whether zero resistivity will be realized around 65 or not, further challenges should be made.
Before our work, fabrication of high-quality FeCh ultrathin films using PLD was believed to be challenging.
As we described in the previous section, the key factor for the reproducible realization of high $T_c$ SC of category 3 is using an atomically flat substrate surface with step-terrace structure.
As mentioned earlier, PLD offers unique advantages, such as its flexible ability to engineer interfaces with a wide variety of materials.
These advantages are promising to design the novel heterostructure to enhance $T_c$ or realizing novel functions, as we worked on FeSe/TiO$_2$ and FeSe/FeTe superlattices.

 So far, we focused on $T_c$ behavior (phase diagram) and related electronic structures. 
 In terms of the future potential of FeCh as a practical superconductor, pinning properties and critical current behavior is most important.
Since right after the discovery of superconductivity in FeCh, tremendous number of studies have been done to investigate the pinning properties and to improve critical current behavior.
These are the issues of independent review article by themselves, and we do not survey these here.
However, FeCh films have several characteristic advantages in terms of pinning properties, partly because of lower anisotropy and less brittleness than, for instance, cuprate high-$T_c$ superconductor.
It keeps high $J_c$ until high magnetic fields instead of relatively low $T_c$.
For example, strongly strained films grown on CaF$_2$ have $J_c$ of $\sim$ 1 MA/cm$^2$ at zero magnetic field and 0.5MA/cm$^2$ at 9 T\cite{Tsukada2011,Braccini2013}.
Films grown on CeO$_2$ buffer layers exhibit $J_c$ higher than 10$^6$ A/cm$^2$ at zero field and  10$^5$ A/cm$^2$ even at 30 T\cite{Si2013}.
Similar insertion of nanoscale composite layers both nonmagnetic\cite{Huang2014} and magnetic\cite{Huang2016} obtained similar numbers for $Jc$.
These clearly demonstrate that as far as the low-temperature high-field operation is concerned FeCh film is a promising candidate which have superior properties than conventional practically used superconductor, NbSe$_3$.

This aspect also must have a merit for device application, where Josephson junction fabrication plays a key role.
Indeed, for example, it has been shown that FeSe$_{0.1}$Te$_{0.9}$ film integrated on Si-based substrate exhibits superconducting properties comparable to those grown on oxide substrates\cite{Huang2018}.

Another characteristic of FeCh is that high $J_c$ of Fe(Se,Te) is realized in topological electronic structures.
This is fascinating in terms of Majorana-based fault-tolerant quantum computing.

Again, PLD is more suitable for these device applications because it is lower in cost and offers a much faster growth rate than MBE. 

In summary, FeCh films, combined with PLD technique, continue to be an extremely fascinating system also from these points of view, in addition to the wide variety of topics that have been discussed in bulk single crystals.

\section*{Acknowledgment}

Dr. Yoshinori Imai (currently at Tohoku University) led the research at the early stages and produced many results summarized in \cite{ImaiR1,ImaiR2,ImaiR3}. He continues to provide valuable discussions with us from time to time.

This research is a long-term project in our laboratory, and many graduate students have contributed significantly to it as co-researchers.
We would like to thank Ryo Ogawa, Yuichi Sawada, Yusuke Kobayashi, Hotaka Kurokawa, Masanori Kawai, Souta Nakamura, Daisuke Asami, Tomoya Ishikawa, Naoya Nakajima, Naoki Shikama, Yuki Sakishita, Jiahui Zhao, Hiroki Ogawa, Kohji Higasa, Hiroki Nakagawa and Gen Matsumoto.

The results presented above are the results of collaborative research with various experts from other institutions and departments. The names of the people involved are (titles omitted, in no particular order): Yuya Kubota, Makina Yabashi, Hiroyuki Ohsumi, Yoshikazu Tanaka, Kenji Tamasaku, Kosuke Nakayama, Takeshi Sato, Phan Ngoc Giao, Takashi Takahashi, Naotaka Yoshikawa, Masayuki Takayama, Ryo Shimano, Masamichi Nakajima, Kazuya Yanase, Setsuko Tajima, Tadashi Adachi, Sang Eun Park, Yuma Kawai, Andreas Suter, Thomas Prokscha, Seiki Komiya, Ataru Ichinose, Yue Sun, Haruhisa Kitano, Shunsuke Kono, Yumiko Katayama, and Kazunori Ueno.

We also thank Profs. Andreas Kreisel, Takami Toyama, Kui Jin, Qihong Chen, and Yihua Wang for their helpful discussions.

This work was supported in part by Grants-in-Aid for Scientific Research (No. 26800187, No. 18K03513, No. 18H04212, No. 18K13500, No. 19K14651, No. JP20H05165), the Murata Science Foundation, and the Precision Measurement Technology Foundation.


\begin{thebibliography}{999}

\bibitem{Hsu}
Hsu F C, Luo J Y, Yeh K W, Chen T K,Huang T W, Wu P M, Lee Y C,
Huang Y L, Chu Y Y Yan D C and Wu M K 2008 {\it Proc. Natl. Acad. Sci.
U.S.A. } \textbf{105} 14262

\bibitem{Mizuguchi}
Mizuguchi Y, Tomioka F, Tsuda S, Yamaguchi T,and Takano Y 2008 {\it Appl. Phys. Lett.} \textbf{93} 152505

\bibitem{Medvedev}
Medvedev S, McQueen T M, Troyan I A, Palasyuk T, Eremets M I, Cava R J, Naghavi S,  Casper F, Ksenofontov V, Wortmann G and Felser C 2009 {\it t. Mater.} \textbf{8} 630

\bibitem{Miyoshi}
Miyoshi K, Morishita K, Mutou E, Kondo M, Seida O, Fujiwara K, Takeuchi J and Nishigori S 2014 {\it J. Phs. Spc. Jpn.} \textbf{83} 013702 

\bibitem{Kothapalli}
Kothapalli K, B\"{o}hmer A E, Jayasekara W T, Ueland B G, Das P, Sapkota A, Taufour V, Xiao Y, Alp E,  Bud'ko, Canfield P C, Kreyssig A and Goldman A I 2016 {\it Nat. Commun.} \textbf{7} 12728 

\bibitem{JPSun}
Sun J P, Matsuura K, Ye G Z, Mizukami Y, Shimozawa M, Matsubayashi K, Yamashita M,
Watashige T, Kasahara S, Matsuda Y, Yan J Q, Sales B C, Uwatoko Y, Cheng J G and Shibauchi T 2016 {\it Nat. Commun.} \textbf{7} 12146 

\bibitem{Guo}
Guo J, Jin S, Wang G, Wang S, Zhu K, Zhou T, He M and Chen X 2010 {\it Phys. Rev. } \textbf{B82}180520(R)

\bibitem{Miyata}
Miyata Y, Nakayama K, Sugawara K, Sato T and Takahashi T 2015 {\it Nat. Mater.} \textbf{14} 775

\bibitem{CHPWen}
Wen C H P, Xu H C, Chen C, Huang Z C, Lou X, Pu Y J, Song Q, Xie B P, Abdel-Hafiez M,
Chareev D A, Vasiliev A N, Peng R andFeng D L 2016 {\it Nat. Commun.} \textbf{7} 10840 

\bibitem{Ying}
Ying T P, Chen X L, Wang G, Jin S F, Zhou T T, Lai X F, Zhang H and Wang W Y 2012 {\it Sci. Rep.} \textbf{2} 426

\bibitem{Lucas}
Burrard-Lucas M, Free D G, Sedlmaier S J, Wright J D, Cassidy S J, Hara Y, Corkett A J, Lancaster T, Baker P J, Blundell S J and  Clarke S J 2013 {\it Nat. Mater.} \textbf{12} 15

\bibitem{Hosono}
Hosono S, Noji T, Hatakeda T, Kawamata T, Kato M and Koike Y 2014 {\it J. Phys. Soc. Jpn.} \textbf{83} 113704

\bibitem{Shiogai}
Shiogai J, Ito Y, Mitsuhashi T, Nojima T and Tsukazaki A 2016 {\it Nat. Phys.} \textbf{12} 42

\bibitem{Hanzawa}
Hanzawa K, Sato H, Hiramatsu H, Kamiya T, Hosono H 2016 {\it Proc. Natl. Acad. Sci.
U.S.A.} \textbf{113} 3986

\bibitem{Lei}
Lei B, Cui J H, Xiang Z J, Shang C, Wang N Z,.Ye G J, Luo X G, Wu T, Sun Z and Chen X H 2016 {\it Phys. Rev. Lett.} \textbf{116} 077002

\bibitem{QYWang}
Wang Q Y, Li Z, Zhang W H, Zhang Z C, Zhang J S, Li W, Ding H, Ou Y B, Deng P and Chang K 2013 {\it Chin. Phys. Lett.} \textbf{29} 037402.

\bibitem{SHe}
He S, He J, Zhang W, Zhao L, Liu D, Liu X, Mou D, Ou Y B, Wang Q Y, Li Z, Wang L, Peng Y, Liu Y, Chen C, Yu L, Liu G, Dong X, Zhang J,  Chen C, Xu Z, Chen X, Ma X, Xue Q and Zhou X J 2013 {\it Nat. Mater.} \textbf{12} 605

\bibitem{XLiu}
Liu X,  Zhao L, He S, He J, Liu D, Mou D, Shen B, Hu Y,
Huang J  and Zhou X J 2015 {\it J. Phys.: Condens. Matter.} \textbf{27} 183201

\bibitem{Bohmer}
B\"{o}hmer A E and Kreisel A 2018 {\it J. Phys.: Condens. Matter.} \textbf{30}
023001

\bibitem{Coldea}
Coldea A I and Watson M D 2018 {\it Annu. Rev. Condens. Matter
Phys.} \textbf{9}, 125

\bibitem{Kreisel}
Kreisel A,  Hirschfeld P J and Andersen B M 2020 {\it Symmetry} \textbf{12}
1402

\bibitem{Shibauchi}
Shibauchi T, Hanaguri T and Matsuda Y 2020 {\it J. Phys. Soc. Jpn.} \textbf{89} 102002


\bibitem{LWang}
Wang L, Ma X and Xue Q K 2016 {\it Supercond. Sci. Technol}. \textbf{29} 123001

\bibitem{ZWang}
Wang A, Liu C, Liu Y and Wang J 2017 {\it J. Phys.: Cond. Mat.} \textbf{29} 153001


\bibitem{DHuang}
Huang D and Hoffman J E 2017 {\it Annu. Rev. Condens. Matter Phys.} \textbf{8} 311

\bibitem{CLiu2020}
Liu C F and Wang J 2020 {\it 2D Mater.} \textbf{7} 022006

\bibitem{SHan2021}
Han S; Song C L, Ma X C and Xue Q K 2021 {\it Comptes. Rendus Physique} \textbf{22 S4} 163

\bibitem{TTanaka2021}
Tanaka T, Ichinokura S, Pedersen A and Hirahara T 2021 {\it Jap. J. Appl. Phys.} \textbf{60} SE0801

\bibitem{DLiu}
Liu D, Zhang W, Mou D, He J, Ou Y B, Wang Q Y,  Li Z, Wang L, Zhao L, He S, Peng Y, Liu X, Chen C, Yu L, Liu G, Dong X, Zhang J, Chen C, XuZ, Hu J, Chen X, Ma X, Xue Q K and Zhou X J 2012 {\it Nat. Commun.} \textbf{3} 931

\bibitem{JLee}
Lee J J, Schmitt F T, Moore R G, Johnston S, Cui Y T, Li W, Yi M, Liu Z K, Hashimoto M, Zhang Y, Lu D H, Devereaux T P, Lee D H and  Shen Z X 2014 {\it Nature} \textbf{515} 245

\bibitem{YCui2015}
Cui Y T, Moore R G, Zhang A M, Tian Y, Lee J J, Schmitt F T, Zhang W H, Li W, Yi M,
Liu Z K, Hashimoto M, Zhang Y, Lu D H, Devereaux T P, Wang L L, Ma X C, Zhang Q M, Xue Q K, Lee D H and Shen Z X 2015 {\it Phys. Rev. Lett.} \textbf{114} 037002

\bibitem{CTang2016}
Tang C J, Liu C, Zhou G, Li F, DingH, Li Z, Zhang D, Li Z, Song C.
Ji S, He K, Wang L, Ma X and Xue Q K 2016 {\it Phys. Rev.} \textbf{B93} 020507(R)

\bibitem{WZhang2016}
Zhang W H, Liu X, Wen C H P, Peng R, Tan S Y, Xie B P, Zhang T and Feng D L 2016 {\it Nano Lett.} \textbf{16} 1969

\bibitem{SZhang2016}
Zhang S Y, Guan J, Jia X, Liu B, Wang W, Li F, Wang L, Ma X, Xue Q K, Zhang J, Plummer E W, Zhu X and Guo J 2016 {\it Phys. Rev.} \textbf{B94} 081116(R)

\bibitem{SRebee2017}
Rebec S N, Jia T, Zhang C, Hashimoto M, Lu D H, Moore R G and Shen Z X 2017 {\it Phys. Rev. Lett.} \textbf{118} 067002


\bibitem{WZhang2017}
Zhang H, Zhang D, Lu X, Liu C, Zhou G, Ma X, Wang L, Jiang P, Xue Q K ansd Bao X 2017 {\it Nat. Commun.} \textbf{8} 214

\bibitem{GZhou2018}
Zhou G, Zhang Q, Zheng F, Zhang D, Liu C, Wang X, Song C L, He K, Ma X C, Gu L, Zhang P, Wang L and Xue Q K 2018 {\it Science Bulletin} \textbf{63} 747

\bibitem{WZhao2018}
Zhao W W, Li M, Chang C Z, Jiang J, Wu L, Liu C, Moodera J S, Zhu Y and Chan M H W 2018 {\it Sci. Adv.} \textbf{4} eaao2682

\bibitem{Saito}
Saito Y, Nojima T and IwasaY 2017 {\it Nat. Rev. Mater.} \textbf{2} 16094

\bibitem{TZhang}
Zhang T, Cheng P, Li W J, Sun Y J, Wang G, Zhu X G, He K, Wang L, Ma X, Chen X, Wang Y, Liu Y, Lin H Q, Jia J F and Xue Q K 2010 {\it Nat. Phys.} \textbf{6} 104

\bibitem{Uchihashi}
Uchihashi T, Mishra P, Aono M  and Nakayama T 2011 {\it Phys. Rev. Lett.} \textbf{107} 207001

\bibitem{MYamada}
Yamada M, Hirahara T and Hasegawa S 2011 {\it Phys. Rev. Lett.} \textbf{110} 237001


\bibitem{Ginzburg}
Ginzburg V 1964 {\it Phys. Lett.} \textbf{13} 101

\bibitem{Bardeen}
Allender D, Bray J and Bardeen J 1973 {\it Phys. Rev.}
\textbf{B7} 1020

\bibitem{BZhao2024}
Zhao D P,  Cui W, Liu Y, Gong G, Zhang L, Jia G, Zang Y, Hu X, Zhang D, Wang Y, Li W, Ji S, Wang L, He K, Ma X andXue Q K 2024 {\it Nat. Commun.} \textbf{15} 3369

\bibitem{WZhang2014}
Zhang W H, Sun Y, Zhang J S, Li F S, Guo M H, Zhao Y F, Zhang H M, Peng J P, Xing Y, Wang H C, Fujita T, Hirata A, Li Z, Ding H, Tang C J, Wang M, Wang Q Y, He K, Ji S H, Chen X, Wang J F, Xia Z C, Li L, Wang Y Y, Wang J, Wang L L, Chen M W, Xue Q K and Ma X C 2014 {\it Chin. Phys. Lett.} \textbf{31} 017401

\bibitem{diamag2}
Zhang Z, Wang Y H, Song Q, Liu C, Peng R, Moler K A, Feng D L and Wang Y 2015 {\it Sci. Bull.} \textbf{60} 1301


\bibitem{GZou2016}
Zhou G; Zhang D; Liu C; Tang C; Wang X; Li Z; Song C; SJi S; He K; Wang L, Ma X and Xue Q K 2016 {\it Appl. Phys. Lett} .\textbf{108} 202603

\bibitem{PZhang2016}
Zhang P, Peng X L, Qian T, Richard P, Shi X, Ma J Z, Fu B B, Guo Y L, Han Z Q,
Wang S C,  Wang L L, Xue Q K,  Hu J P, Sun Y J and Ding H 2018 {\it Phys. Rev.} \textbf{B94} 104510

\bibitem{HDing2016}
Ding H, Lv Y F, Zhao K, Wang W L, Wang L, Song C L, Chen X, Ma X C and Xue Q K 2016 {\it Phys. Rev. Lett.} \textbf{117} 067001

\bibitem{RPeng2014}
Peng R, Xu H C, Tan S Y, Cao H Y, Xia M, Shen X P, Huang Z C, Wen C H P, Song Q, Zhang T, Xie B P, Gong X G and Feng D L 2014 {\it Nat. Commun.} \textbf{5} 5044

\bibitem{HYang2019}
Yang H H, Zhou G, Zhu Y, Gong G M, Zhang G G, Liao M H, Li Z, Ding C, Meng F, Rafique M, Wang H, Gu L, Zhang D, Wang L and Xue Q K 2019 {\it Science Bulletin} \textbf{64} 490

\bibitem{YSong2021}
Song Y, Chen Z, Zhang Q, Xu H, Lou X, Chen X, Xu X, Zhu X, Tao R, Yu T, Ru H, Wang Y, Zhang T, Guo J, Gu L, Xie Y, Peng R and Feng D L 2021 {\it Nat. Commun.} \textbf{12} 5926

\bibitem{CLiu2021}
Liu C, Shin H, Doll A, Kung H H, Day R P, Davidson B A, Dreiser J, Levy G, Damascelli A, Piamonteze C and Zou K 2021 {\it npj Quantum Mat.} \textbf{6} 85

\bibitem{CLi2024}
Li C, Song Y, Wang X, Lei M, Chen X, Xu H, Peng R and Feng D L 2024 {\it Nano Lett.} \textbf{24} 8303

\bibitem{GZhou2021}
Zhou G Y, Zhang Q H, Zheng F, Zhang D, Liu C, Wang X X, Song C L, He K, Ma X C, Gu L, Zhang P, Wang L and Xue Q K 2018 {\it Science Bulletin} \textbf{63} 747


\bibitem{Biswas2018}
Biswas P K, Salman Z, Song Q, Peng R, Zhang J, Shu L,  Feng D L, Prokscha T and Morenzon E 2018 {\it Phys. Rev.} \textbf{B97} 174509



\bibitem{Pedersen}
Pedersen A K, Ichinokura S, Tanaka T, Shimizu R, Hitosugi T and Hirahara T 2020 {\it Phys. Rev. Lett.} \textbf{124} 227002

\bibitem{Faeth}
Faeth B D, Yang S L, Kawasaki J K, Nelson J N. Mishra P, Parzyck C T, Li C, Schlom D G and Shen K M 2021 {\it Phys. Rev.} \textbf{X11} 021504

\bibitem{Ge}
Ge J F, Liu Z L, Liu C, Gao C L, Qian D, Xue Q K, Liu Y and Jia J F 2015 {\it Nat. Mater.} \textbf{14} (2015) 285

\bibitem{Bozovic}
Bozovic I and Ahn C 2014 {\it Nat. Phys.} \textbf{10} 892


\bibitem{Sprau}
Sprau P O, Kostin A, Kreisel A, B\"{o}hmer A E, Taufour V, Canfield P C, Mukherjee S, Hirschfeld P J, Andersen B M and Davis J C S 2017 {\it Science} \textbf{357} 75


\bibitem{BLei}
Lei B, Cui J H, Xiang Z J, Shang C, Wang N Z, Ye G J,Luo X G, Wu T, Sun Z and Chen X H 2016 {\it Phys. Rev. Lett.} \textbf{116} 077002

\bibitem{Farrar}
Farrar L S, Bristow M, Haghighirad A A, McCollam A, Bending S J and Coldea A I 2020 {\it npj Quantum Mater.} \textbf{5} 29

\bibitem{Xie}
Xie J, Liu X, Zhang W, Wong S M, Zhou X F, Zhao Y S, Wang S M, Lai K T and Goh S K 2021 {\it Nano Lett.} \textbf{21} 9310

\bibitem{Zhu}
Zhu C S, Lei B, Sun Z L, Cui J H, Shi M Z, Zhuo W Z, Luo X G and Chen X H 2021 {\it Phys. Rev.} \textbf{B104} 024509

\bibitem{BKT}
Berezinskii V L 1970 {\it Sov. Phys. JETP} \textbf{32} 493

\bibitem{BKT2}
Kosterlitz J M and Thouless D J 1973 {\it J. Phys.} \textbf{C6} 1181

\bibitem{YLi}
Li Y, Liu H, Ji H, Ji C C, Qi S, Jiao X, Dong W F, Sun Y, Zhang W H,
Cui Z, Pan M H, Samarth N, Wang L, Xie X C, Xue Q K, Liu Y and Wang J
 2024 {\it Phys. Rev. Lett.} \textbf{132} 226003


\bibitem{Shikama1}
Shikama N, Sakishita Y, Nabeshima F, Katayama Y, Ueno K and Maeda A 2020 {\it Appl. Phys. Express} \textbf{13} 083006

\bibitem{Shikama2}
Shikama N, Sakishita Y,. Nabeshima F and Maeda A 2021 {\it Phys. Rev.} \textbf{B104} 094512



\bibitem{Hiramatsu}
Hiramatsu H, Katase T, Kamiya T and Hosono H 2012 {\it J. Phys. Soc. Jpn.} \textbf{81} 011011

\bibitem{Haindl}
Haindl S, Kidszun M, Oswald S, Hess C, B\"{o}chner B, K\"{o}lling S, Wilde L, Thersleff T, Yurchenko V V, Jourdan M,  Hiramatsu H and Hosono H 2014 {\it Rep. Prog. Phys.} \textbf{77} 046502


\bibitem{Imai1}
Imai Y, Sawada Y, Nabeshima F and Maeda A 2015 {\it Proc. Natl. Acad. Sci. USA} \textbf{112} 1937

\bibitem{Imai2}
Imai Y, Sawada Y, Nabeshima Y, Asami D, Kawai M and Maeda A 2017 {\it Sci. Rep.} \textbf{7} 46653

\bibitem{JZhuang}
Zhuang J, Yeoh W K, Cui X Y, Xu X, Du Y, Shi Z X, Ringer S P, Wang X L and Dou S X 2014 {\it Sci. Rep} \textbf{4} 7273 

\bibitem{Nabeshima1}
Nabeshima F, Ishikawa T, Oyanagi K, Kawai M and Maeda A 2018 {\it J. Phys. Soc. Jpn.} \textbf{87} 073704


\bibitem{ImaiR1}
Imai Y, Nabeshima F and Maeda A 2015 {\it Solid State Physics} {\bf 50} 457. ({\it in Japanese})

\bibitem{ImaiR2}
Imai Y, Nabeshima F and Maeda A, 
"Comparative Review on Thin Film Growth of Iron-Based Superconductors" 2017 {\it Condens. Matter} \textbf{2} 25/1-33 ({\it in Japansese}).

\bibitem{ImaiR3}
Imai Y, Nabeshima F and Maeda A 2018 {\it Applied Physics} \textbf{87} 926 ({\it in Japanese}).


\bibitem{Nabeshima2}
Nabeshima F, Kawai M, Ishikawa T, Shikama N and Maeda A 2018 {\it Jpn. J. Appl. Phys.} \textbf{57} 120314

\bibitem{Huang2025}
Huang Z, Wang X, Shi Q, Zhao Z, Yuan C,
Wang Y, Jiang Y, Jin Y, Zhu B, Yuan J,  Kusmartsev F, Maeda A, Lin Z,  Chen Q and Jin K {\it Chin. Phys. Lett.} {\it submitted.}

\bibitem{ZFeng2024}
Feng Z, Zhang H, Yuan J, Jiang X, Wu X, Zhao Z, Xu Q, Stanev V, Zhang Q, Yang H, Gu L, Meng S, Weng S, Chen Q, Takeuchi I, Jin K and Zhao Z 2024 {\it Qunatum Frontiers} \textbf{3} 12

\bibitem{Ghini}
Ghini M, Bristow M, Prentice J C A,, Sutherland S, Sanna S, Haghighirad A A  and Coldea A I 2021 {\it Phys. Rev.} \textbf{B103} 205139

\bibitem{Nakajima1}
Nakajima M, Ohata Y and Tajima S 2021 {\it Phys. Rev. Mater.} \textbf{5} 044801

\bibitem{Phan}
Phan G N, Nakayama K, Sugawara K, Sato T, Urata T, Tanabe Y, Tanigaki K, Nabeshima F, Imai Y, Maeda A and Takahashi T 2017 {\it  Phys. Rev.}\textbf{B95} 224507

\bibitem{Fang}
Fang M H, Pham H M,, Qian B, Liu T J, Vehstedt E K,
Liu Y, Spinu L and Mao Z Q 2008 {\it Phys. Rev.} \textbf{B78}, 224503

\bibitem{Mizuguchi2}
Mizuguchi Y, Tomioka F, Tsuda S, Yamaguchi T and Takano Y 2009 {\it J. Phys. Soc. Jpn.} \textbf{78} 074712

\bibitem{Watson1}
Watson M D, Kim T K,  Haghighirad A A, Blake S F, Davies N R, Hoesch M, Wolf T and Coldea A I 2015 {\it Phys. Rev.} \textbf{B92}
121108(R)

\bibitem{bulkSchange1}
Mizukami Y, Haze M, Tanaka O, Matsuura K, Sano D, B\"{o}ker J, Eremin I, Kasahara S, Matsuda Y and Shibauchi T 2023 {\it  Commun. Phys.} \textbf{6} 183


\bibitem{bulkSchange2}
Sato Y, Kasahara S, Taniguchi T, Xing X, Kasahara Y, Tokiwa Y, Yamakawa Y, Kontani H, Shibauchi T and Matsuda Y 2018 {\it Proc. Nat. Acad. Sci.} \textbf{115} 1227


\bibitem{Shi_FeSeS}
Yi X, Xing X, Qin L, Feng J, Li M, Zhang Y, Meng Y, Zhou N, Sun Y and Shi Z 2021 {\it Phys. Rev.} \textbf{B103} 144501


\bibitem{Terao}
Terao K, Kashiwagi T, Shizu T, Klemm R A  and Kadowaki K 2019 {\it Phys. Rev.} \textbf{B100} 224516


\bibitem{SLiu}
Liu S, Yuan J, Huh S, Ryu H, Ma M, Hu W, Li D, Ma S, Ni S, Shen P, Jin , Yu L, Kim C, Zhou F, Dong X, Zhao Z 2009 {\it arXiv. 2009.13286}


\bibitem{Mukasa}
Mukasa K, Matsuura K, Qiu M, Saito M, Sugimura Y,
Ishida K, Otani M, Onishi Y, Mizukami Y, Hashimoto K,
Gouchi J, Kumai R, Uwatoko Y and Shibauchi T 2021 {\it Nat.
Commun.} \textbf{12} 381



\bibitem{bulkQCP}
Ishida K, Onishi Y, Tsuji M, Mukasa K, Qiu M, Saito M, Sugimura Y, Matsuura K, Mizukami Y, Hashimoto K and Shibauchi T 2022 {\it Proc. Nat. Acad, Sci.} \textbf{119} e2110501119



\bibitem{Qihong}
Lin Z F, Tu S, Xu J, Shi Y, Zhu B, Dong C, Yuan J, Dong X, Chen Q, Li Y,Jin K and Zhao Z 2022 {\it Sci. Bull.} \textbf{67} 1443

\bibitem{Sato2024}
Sato Y, Nagahama S, Belopolski I, Yoshimi R, Kawamura M,, Tsukazaki A,, Kanazawa N, Takahashi K S, Kawasaki M and Tokura Y 2024 {\it Phys. Rev. Mat.} \textbf{8} L041801

\bibitem{Sato2025a}
Sato Y, Nagahama S, Belopolski I, Yoshimi R, Kawamura M, Tsukazaki A, Yamada A, Tokunaga M, Kanazawa N, Takahashi K S,
Onuki Y, Kawasaki M and Tokura Y 2025 {\it Phys. Rev.} \textbf{B112} L041121


\bibitem{Yi2024}
Yi H, Zhao Y F, Chan Y T, Cai J, Mei R, Wu X, Yan Z J, Zhou L J, Zhang R, Wang Z, Paolini S, Xiao R, Wang K, Richardella A R, Singleton J, Winter L E, Prokscha T, Salman Z, Suter A, Balakrishnan P P, Grutter A J, Chan M H W, Samarth N, Xu X D, Wu W, Liu C X and Chang C Z 2024 {\it Science} \textbf{383} 634

\bibitem{Sato2025b}
Sato Y, Nagahama S, Kitou S,, Sagayama H, Belopolski I, Yoshimi R, Kawamura M,, Tsukazaki A,  Kanazawa N, Nomoto T, Arita R,
Arima T,. Kawasaki M and Tokura Y 2025 {\it Nat. Commun.} \textbf{16} 10913 


\bibitem{Kubota}
Kubota Y, Nabeshima F,. Nakayama K, Ohsumi H, Tanaka Y, Tamasaku K, Suzuki T,  Okazaki K,. Sato T, Maeda A and Yabashi M 2023 {\it Phys. Rev.} \textbf{B108} (2023) L100501

\bibitem{ortho}
Margadonna S, Takabayashi Y, McDonald M T, Kasperkiewicz K, Mizuguchi Y, Takano Y, Fitch A N, Suard E and Prassides K 2008 {\it Chem. Commun.} 5607

\bibitem{nematic}
Fernandes R M, Chubukov A V and Schmalian J 2014 {\it Nat. Phys.} \textbf{10} 97

\bibitem{Nakayama1}
Nakayama K,.Tsubono R, Phan G N, Nabeshima F, Shikama N, Ishikawa T, Sakishita Y, Ideta S, Tanaka K, Maeda A, Takahashi T and Sato T 2021 {\it Phys. Rev. Res.} \textbf{3} L012007

\bibitem{bulk_ARPES}
Nakayama K, Miyata Y, Phan G N, Sato T, Tanabe Y, Urata T, Tanigaki K and Takahashi T 2014 {\it Phys. Rev. Lett.} \textbf{113} 237001

\bibitem{Nabeshima3}
Nabeshima F, Ishikawa T, Shikama N and Maeda A 2020 {\it Phys. Rev.} \textbf{B101} 184517

\bibitem{Sawada}
Sawada Y, Nabeshima F, Imai Y and Maeda A 2016 {\it J. Phys. Soc. Jpn.} \textbf{85} 073703

\bibitem{Terashima}
Terashima T, Kikugawa N, Kiswandhi A, Choi E S, Brooks J S, Kasahara S, Watashige T, Ikeda H, Shibauchi T,
Matsuda Y, Wolf T, B\"{o}hmer A E, Hardy F, Meingast C, L\"{o}hneysen H v, Suzuki M, Arita R and Uji S 2014 {\it Phys. Rev.} \textbf{B90} 144517


\bibitem{Yoshikawa}
Yoshikawa N, Takayama M, Shikama N, Ishikawa T, Nabeshima F, Maeda A and Shimano R 2019 {\it Phys. Rev.} \textbf{B100} 035110


\bibitem{Nakajima2}
Nakajima M, Yanase K Nabeshima F, Imai Y, Maeda A and Tajima S 2017 {\it Phys. Rev.} \textbf{B95} 184502

\bibitem{Nakajima3}
Nakajima M, Yanase K Kawai M, Asami D, IshikawaT, Nabeshima F, Imai Y, Maeda A and Tajima S 2021 {\it Phys. Rev.} \textbf{B104} 024512

\bibitem{FBSoptical1}
Nakajima M, Ishida S,. Kihou K, Tomioka Y, Ito T, Yoshida Y,  Lee C H, Kito H,  Iyo A,
Eisaki H, KojimaK M and Uchida S 2010 {\it Phys. Rev.} \textbf{B81} 104528

\bibitem{FBSoptical2}
Homes C C, Dai Y M, Wen J S,  Xu Z J and Gu G D 2015 {\it Phys. Rev.} \textbf{B91} 144503

\bibitem{FBSoptical3}
Wu D, Bari\v{s}i\'{c} N, Kallina P, Faridian A, Gorshunov B, Drichko N, Li N J, Lin X, Cao G H,
Xu Z A, Wang N L and Dressel M 2010 {\it Phys. Rev.} \textbf{B81} 100512(R)

\bibitem{FBSoptical4}
Nakajima M, Ishida S, Tanaka T, Kihou K, Tomioka Y, Saito T, Lee C H, Fukazawa H, Kohori Y, Kakeshita T, Iyo A, Ito T, Eisaki H and Uchida S 2014 {\it Sci. Rep.} \textbf{4} 5873




\bibitem{Kurokawa}
Kurokawa H, Nakamura S, Zhao J, Shikama N, Sakishita Y,Sun Y, Nabeshima F,
Imai, Y, Kitano H and Maeda A 2021 {\it Phys. Rev.} \textbf{B104} 014505

\bibitem{XLong}
Long X, Zhang S, Wang F and Liu Z 2020 {\it npj Quantum Mater.} \textbf{5} 50

\bibitem{Yamada}
Yamada T and Tohyama T 2021 {\it Phys. Rev.} \textbf{B104} L161110


\bibitem{Nabeshima4}
Nabeshima F, Kawai Y, Shikama N, Sakishita Y,Suter A, Prokscha T, Park S. E, Komiya S,
Ichinose A,,. Adachi T and Maeda A 2021 {\it Phys. Rev.} \textbf{B103} 184504


\bibitem{Adachinew}
Miwa K, Kawa Y, Nabeshima F, Suter A, Salman Z, Prokscha T,  Maeda A and Adachi T 2026 {\it Proc. 38th Int. Symp. on Superconductivity, Nagasaki, Japan} {\it submitted}

\bibitem{Takigawa}
Imai T, Ahilan K, Ning F L, McQueen T M and Cava R J 2009 {\it Phys. Rev. Lett.} \textbf{102} 177005





\bibitem{CLSong}
Song,C L, Wang Y L, Cheng P, Jiang Y P, Li W, Zhang T, Li Z,
He K, Wang L, Jia J F, Hung H H3 Wu C, Ma X, Chen X and Xue Q K 2011 {\it Science} \textbf{332} 1412


\bibitem{Hanaguri1}
Hanaguri T, Niitaka S, Kuroki K and Takagi H 2010 {\it Science} \textbf{328} 474



\bibitem{Hanaguri2}
Hanaguri T, Iwaya K, Kohsaka Y, Machida T, Watashige T, Kasahara S, Shibauchi T and Matsuda Y 2018 {\it Sci. Adv.} \textbf{25} eaar6419



\bibitem{method}
{\it For example}, Maeda A, Kitano H and Inoue R 2006 {\it J. Phys. Cond. Matt.} \textbf{17} R143


\bibitem{Barannik}
Barannik A A, Cherpak N T, Kharchenko M S, Wu Y, Luo S, He Y and Porch A 2014 {\it Low Temp. Phys.} \textbf{40} 492

\bibitem{Klein}
Klein N, Chaloupka H, M\"{u}ller G, Orbach S, Piel H, Roas B,
Schultz L, Klein U and Peiniger M 1990 {\it J. Appl. Phys.} \textbf{67} 6940

\bibitem{Drabeck}
Drabeck L, Holczer K,  Gr\"{u}ner G, Chang J J, Scalapino D J,
 Inam A, Wu X D, Nazar L and Venkatesan T1990 {\it Phys. Rev.}
\textbf{B42} 10020

\bibitem{Schaumberg}
Schaumburg G and Helberg H W 1992 {\it Ann. Physik} \textbf{1} 584


\bibitem{Peli1}
Peligrad D N, Nebendahl B, Kessler C, Mehring M, Dulcic A, Po\v{z}ek M and Paar D 1998 {\it Phys. Rev.} \textbf{B58} 11652

\bibitem{Peli2}
Peligrad D N, Nebendahl B, Mehring M,
Dul\v{c}i\'{c} A, Po\v{z}ek M and Paar D 2001 {\it Phys. Rev.} \textbf{B64} 224504

\bibitem{Matsumoto}
Matsumoto G, Ogawa R, Higasa K, Kobayashi T, Nakagawa H and Maeda A 2024 {\it J. Phys. Conf. Ser.} \textbf{2776} 012002, also {\it arXiv.2402.18082}


\bibitem{Takahashi}
Takahashi H, Imai Y, Komiya S, Tsukada I and Maeda A 2011 {\it Phys. Rev.} \textbf{B84} 132503


\bibitem{Mishra}
Mishra V, Graser S and Hirschfeld P J 2011 {\it Phys. Rev.} \textbf{B84} 014524

\bibitem{Prozorov}
Prozorov R and Kogan V G 2011 {\it Rep. Prog. Phys.} \textbf{74} 124505

\bibitem{Uemura}
Uemura Y J, Luke G M, Sternlieb B J, Brewer J H, Carolan J F, Hardy W N, Kadono R, Kempton J R, Kiefl R F,
Kreitzman S R, Mulhern P, Riseman T M, Williams D L, Yang B X, Uchida S, Takagi H, Gopalakrishnan J, Sleight A W, Subramanian M,  Chien C L, Cieplak M Z, Xiao G, Lee V Y, Statt B W, Stronach C E, Kossler W J and Yu X H 1989 {\it Phys. Rev. Lett.} \textbf{62} 2317

\bibitem{Uemura2}
Uemura Y J 1997 {\it Physica} \textbf{C282-287} 194



\bibitem{KYLiang2025}
Liang K Y, Zhang R Z, Lin Z F, Li Z J, Chen B R, Zhang P H, Yao K Z, He Q S, Zhou Q Z, Yao X H, Jin K and Wang Y H {\it arXiv. 2505.16184v1}



\bibitem{Bonn}
Bonn D A, Dosanjh P, Liang R and Hardy W N 1992 {\it Phys. Rev.
Lett.} \textbf{68} 2390

\bibitem{Okada}
Okada T, Imai Y, Urata T, Tanabe Y, Tanigaki K and Maeda A 2021 {\it J. Phys. Soc. Jpn.} \textbf{90} 094704

\bibitem{MLi}
Li M,  Lee-Hone N R, Chi S, Liang R, Hardy W N, Bonn D A, Girt E and Broun D M 2018 {\it New J. Phys.} \textbf{18} 082001


\bibitem{Hirschfeld}
Hirschfeld P J, Putikka W O and Scalapino D J 1993 {\it  Phys. Rev.Lett.} \textbf{71} 3705

\bibitem{Ozcan}
\"{O}zean S, Turner P J, Waldram J R, Drost R J, Kes P H and Broun D M 2006 {\it Phys. Rev.} \textbf{B73} 064506

\bibitem{Hashimoto}
Hashimoto K, Shibauchi T, Kasahara S, Ikada K, Tonegawa S, Kato T, Okazaki R, van der Beek C J and Konczykowski M 2009 {\it Phys. Rev. Lett.} \textbf{102} 207001

\bibitem{Quinlan}
Quinlan S M, Scalapino D J and Bulut N 1994 {\it Phys. Rev.} \textbf{B49} 1470



\bibitem{Labat2017}
Labat D and Paul L 2017 {\it Phys. Rev.} \textbf{B 96} 195146


\bibitem{Lubashevsky2012}
Lubashevsky Y, Lahoud E, Chashka K, Podolsky D and Kanigel A 2012 {\it Nat. Phys.} \textbf{8} 309

\bibitem{Okazaki2014}
Okazaki K, Ito Y, Ota Y, Kotani Y, Shimojima T, Kiss T, Watanabe S, Chen C T, Niitaka S, Hanaguri T, Takagi H, Chainani A andShin S 2014 {\it Sci. Rep.} \textbf{4} 4013

\bibitem{Kasahara2014}
Kasahara S, Watashige T, Hanaguri T, Kohsaka Y, Yamashita T. Shimoyama Y, Mizukami Y, Endo R, Ikeda H, Aoyama K, Terashima T, Uji S, Wolf, Lohneysen H V, Shibauchi T, Matsuda Y 2014 {\it Proc. Nat. Aad. Sci.} \textbf{111} 16309

\bibitem{LarkinBK}
Larkin A I and Bariamov A A 2003 {\it Fluctuation Phenomena in Superconductors} in {\it The Physics of Superconductors (eds. Bennemann K H and Ketterson J B)} vol. 1 Berlin-Heidelberk-New York, Springer p 95

\bibitem{Kasahara2016}
Kasahara S, Yamashita T, Shi A, Kobayashi R, Shimoyama Y, Watashige T, Ishida K, Terashima T, Wolf T, Hardy F, Meingast C, v. L\"{o}hneysen H, Levchenko A, Shibauchi T and Matsuda Y 2016 {\it Nat. Commun.} \textbf{7} 128433

\bibitem{Takahashi2019}
Takahashi H, Nabeshima F, Ogawa R, Ohmichi E, Ohta H and Maeda A 2019 {\it Phys. Rev.} \textbf{B99} 060503(R)

\bibitem{Yuan2017}
Yuan H, Chen G, Zhu X, Xing J and Wen H H 2017 {\it Phys. Rev.} \textbf{B96} 064501

\bibitem{Shi2018}
Shi A, Arai T, Kitagawa S, Yamanaka T, Ishida K, B\"{o}hmer A E, Meingast C, Wolf T, Hirata M and Sasaki T 2018 {\it J. Phys. Soc. Jpn.} \textbf{87} 013704

\bibitem{Booth}
Booth J C, Wu D H and Anlage S M 1994 {\it Rev. Sci. Inst.} \textbf{65} 2082

\bibitem{KitanoRSI}
Kitano H, Ohashi T and Maeda A 2008 {\it Rev. Sci. Inst.} \textbf{79} 074701

\bibitem{Nabeshimafluc}
Nabeshima F, Nagasawa K and Maeda A 2018 {\it Phys. Rev.} \textbf{B97} 024504

\bibitem{Isoyama2021}
Isoyama K, Yoshikawa N, Katsumi K, Wong J, Shikama N, Sakishita Y, Nabeshima F, Maeda A and Shimano R 2021 {\it Commun. Phys.} \textbf{4} 160









\bibitem{Ueno}
Ueno K, Nakamura S, Shimotani H,  Ohtomo A, Kimura N, Nojima T, Aoki H, Iwasa Y and Kawasaki M 2008 {\it Nat. Mater.} \textbf{7} 855

\bibitem{Shiogai2}
Shiogai J, Miyakawa T, Ito Y, Nojima T and Tsukazaki A 2017 {\it Phys. Rev.} \textbf{B 95} 115101

\bibitem{Kouno}
Kouno S, Sato Y, Katayama Y, Ichinose A, Asami D, Nabeshima F,  Imai Y, Maeda A and Ueno K 2018 {\it Sci. Rep.} \textbf{8} 14731

\bibitem{Maekawa}
Maekawa S and Fukuyama H 1982 {\it J. Phys. Soc. Jpn.} \textbf{51} 1380




\bibitem{Jiang2023}
Jiang X, Qin M, Wei X, Xu L, Ke J, Zhu H, Zhang R, Zhao Z, Liang Q, Wei Z, Lin Z, Feng Z, Chen F, Xiong P, Yuan J, Zhu B, Li Y, Xi C, Wang Z, Yang M, Wang J, Xiang T, Hu J, Jiang K, Chen Q, Jin K and Zhao Z 2023 {\it Nat. Phys.} \textbf{19} 365




\bibitem{Zhang2025}
Zhang R, Qin M, Li C, Zhao Z, Wei Z, Xu J, Jiang X, Cheng W, Shi Q, Wang X, Yuan J, Li Y, Chen Q, Xiang T, Sachdev S, Li Z X, Jin K and Zhao Z 2025 {\it Sci. Adv.} \textbf{11} eadu0795

\bibitem{LASER-MBE}
Koinuma H, Nagata H, Tsukahara T, Gonda S and Yoshimoto M 1991 {\it Appl. Phys. Lett.} \textbf{58} 2027

\bibitem{Pascal}
{\it For example}, http://www.pascal-co-ltd.co.jp/product/deppld.html.

\bibitem{Shiogai3}
Shiogai J, Kimura S, Awaji S, Nojima T and Tsukazaki A 2018 {\it Phys. Rev.} \textbf{B97} 174520

\bibitem{Tanaka0}
Tanaka T, Akiyama K, Yoshino R and Hirahar T 2018 {\it Phys. Rev.} textbf{B98} 121410R

\bibitem{Tanaka1}
Tanaka T, Akiyama K, Ichinokura S, Shimizu R, Hitosugi T and Hirahara T 2020 {\it Phys. Rev.} \textbf{B101} 205421

\bibitem{Tanaka2}
Tanaka T, Ichinokura S, Pedersen A and Hirahara T 2021 {\it Jap. J. Appl. Phys.} \textbf{60} SE0801


\bibitem{Kobayashi}
Kobayashi T, Ogawa H, Nabeshima F and Maeda A 2022 {\it Supercond. Sci. Technol.} \textbf{35} 07LT01

\bibitem{Wang2}
Wang Q, Zhang W, Zhang Z, Sun Y, Xing Y, Wang Y, Wang L, Ma X, Xue Q K and Wang J 2015 {\it 2D Mater.} \textbf{2} 044012

\bibitem{Kobayashi2}
Kobayashi T, Nakagawa H, Ogawa H, Nabeshima F and Maeda A 2023 {\it J. Phys. Condens. Matter}. \textbf{35} 41LT01

\bibitem{Wang3}
Fan Q, Zhang W H  Liu X, Yan Y J, Ren M Q, Peng R, Xu H C, Xie B P, Hu J P, Zhang T and Feng D L 2015 {\it Nat. Phys.} \textbf{11} 946

\bibitem{Zhao2024}
Zhao Z, Wei Z, Yu X, Dong W, Huang J, Chen C, Feng Z, Qin M, Wang X,
Cao S,, Huan Q, Li Y, Yuan J, Zhu B, Chen Q, Sun Y, Wang L, Qian T and Jin K 2024 {\it Phys. Rev.} \textbf{B110} L140507

\bibitem{Kobayashi3}
Kobayashi T, Ogawa R and Maeda A 2025 {\it Phys. Rev.} \textbf{B 112} 094525



\bibitem{diamag1}
Deng L Z, Lv B, Wu Z, Xue Y Y, Zhang W H, Li F S, Wang L L, Ma X C, Xue Q K and Chu C. W 2014 {\it Phys. Rev.} \textbf{B90} 214513


\bibitem{YSun2014}
Sun Y, Zhang W, Xing Y, Li F, Zhao Y, Xia Z, Wang L, Ma X, Xue Q K and Wang J 2014 {\it Sci. Rep.} \textbf{4} 6040




\bibitem{MPA}
Fisher M P A 1994 {\it Phys. Rev. Lett.} \textbf{65} 923

\bibitem{SIT}
Haviland D B, Liu Y and Goldman A M 1989 {\it Phys. Rev. Lett.} \textbf{62} 2180

\bibitem{SITr}
Kapitulnik A, Kivelson S A and Spivak B 2019 {\it Rev. Mod. Phys.} \textbf{91} 11002

\bibitem{Schneider}
Schneider R, Zaitsev A G, Fuchs D and L\"{o}hneysen H V 2012 {\it Phys. Rev. Lett.} \textbf{108} 257003

\bibitem{Ienaga}
Ienaga K, Hayashi T, Tamoto Y, Kaneko S and Okuma S 2020 {\it Phys. Rev. Lett.} \textbf{125} 257001


\bibitem{Swanson}
Swanson M, Loh Y L, Randeria M and Trivedi N 2014 {\it Phys. Rev.} \textbf{X4} 021007


\bibitem{NabeshimaSL}
Nabeshima F, Imai Y, Ichinose a, Tsukada I and MaedaA 2017 {\it Jpn. J. Appl. Phys.} \textbf{56} 020308

\bibitem{KobayashiNY}
Kobayashi T, Nakagawa H, Ogawa H and Maeda A {\it arXiv. 2509.23591}



\bibitem{Tsukada2011}
Tsukada I, Hanawa M, Akiike T, Nabeshima F, Imai Y, Ichinose A, Komiya S, Hikage T, Kawaguchi T, Ikuta H and Maeda A, 2011 {\it Appl. Phys.Express} \textbf{4} 053101


\bibitem{Braccini2013}
Braccini V; Kawale S; Reich E; Bellingeri E; Pellegrino L; Sala A; Putti M; Higashikawa K; Kiss T;
Holzapfel B and Ferdeghini C 2013 {\it Appl. Phys. Lett.} \textbf{103} 172601


\bibitem{Si2013}
Si W, Han S J, Shi X, Ehrlich S N, Jaroszynski,  Goyal A and Li Q 2013
{\it Nature Commun.} \textbf{4} 1347


\bibitem{Huang2014}
Huang J,, Chen L, Jian J, Khatkhatay F and Wang H 2014
{\it Supercond. Sci. Technol.} \textbf{27} 105006


\bibitem{Huang2016}
Huang J, Chen L, Jian J, Tyler K, Li L, Wang H and Wang H 2016
{\it J. Phys.: Condens. Matter} \textbf{28} 025702


\bibitem{Huang2018}
Huang J, Chen L, Li L, Qi z, Sun X,
Zhang X and Wang H 2018 {\it J. Phys.} {\bf D}: {\it Appl. Phys.} \textbf{51} 205301



\end{thebibliography}
\end{document}